\documentclass[useAMS,usenatbib]{mn2e}

\usepackage{graphicx}

\title[Faint Spectrophotometric Standards]{New Faint Optical Spectrophotometric Standards. Hot White Dwarfs from the Sloan Digital Sky Survey}
\author[C. Allende Prieto, I. Hubeny and J. A. Smith]{Carlos Allende Prieto$^{1}$\thanks{E-mail:
cap@mssl.ucl.ac.uk},
Ivan Hubeny$^{2}$ and
J. Allyn Smith$^{3}$
\\
$^{1}$Mullard Space Science Laboratory, University College London, Holmbury St Mary, Dorking, Surrey RH5 6NT, UK\\
$^{2}$Department of Astronomy and Steward Observatory, University of Arizona, Tucson, AZ 85721, USA \\
$^{3}$Department of Physics and Astronomy, Austin Peay State University, P.O. 
Box 4608, Clarksville, TN 37044, USA}
\begin{document}

\date{To appear in MNRAS; submitted on February 4, 2009; accepted on March  12, 2009}

\pagerange{\pageref{firstpage}--\pageref{lastpage}} \pubyear{2002}

\maketitle

\label{firstpage}

\begin{abstract}
The spectral energy distributions for pure-hydrogen (DA) hot white dwarfs can
be accurately predicted by model atmospheres. This makes it possible to define 
spectrophotometric calibrators by scaling the theoretical spectral
shapes with broad-band photometric observations  
-- a strategy successfully exploited for the
spectrographs onboard the Hubble Space Telescope (HST) 
using three primary DA standards.
Absolute fluxes for non-DA secondary standards, introduced to 
increase the density of calibrators in the sky, need to be
referred to the primary standards, but a far better solution
would be to  employ a network of DA stars scattered throughout the sky. 
We search for blue objects in the
sixth data release of the Sloan Digital Sky Survey (SDSS) and fit DA model fluxes to 
identify suitable candidates. Reddening needs to be considered in the analysis
of the hottest and therefore more distant stars.
We propose a list of nine pure-hydrogen white dwarfs 
with absolute fluxes with estimated uncertainties below 3 \%, 
including four objects with
estimated errors $<2$ \%, as candidates for spectrophotometric standards 
in the range $14<g<18$, and provide   
model-based fluxes scaled to match the SDSS broad-band fluxes for each.
We apply the same method to the three HST DA standards, linking the zero point
of their absolute fluxes to $ugr$ magnitudes transformed from photometry obtained
with the USNO 1-m telescope. 
For these stars we estimate uncertainties of $<1$ \% in the optical, finding good consistency 
with the fluxes adopted for HST calibration.
\end{abstract}


\section{Introduction}

Good flux standards are hard to get. Ground-based observations are limited
in accuracy by time-dependent variations in transparency of the
terrestrial atmosphere. Spaced-based fluxes are free from atmospheric 
distortions, but are more difficult to relate to standard sources in 
the laboratory. In addition, charge-transfer efficiency problems 
in CCDs become more acute in space due to bombardment by
high-energy particles.
Despite these issues, the Hubble Space Telescope 
CALSPEC spectrophotometric standards, 
advertised to provide absolute fluxes good to about 1-2 \% in the 
optical and near-infrared, constitute the preferred reference.

The CALSPEC absolute fluxes are based on model atmospheres for three hot 
DA white dwarfs normalized to Landolt $V$-band photometry  (Bohlin 1996, 2000; 
Bohlin, Dickinson \& Calzetti 2001). In this calibration, the spectral energy 
distribution postulated for these  stars is computed with hydrostatic 
plane-parallel NLTE pure-hydrogen model atmospheres calculated with the
code Tlusty (see \S \ref{analysis}). These atmospheres
consist of pure hydrogen, and are free from convection. 
Under the assumption that the physics of DA atmospheres
is well-known, these objects are then promoted to space calibration sources
(see, e.g., Holberg et al. 1982, 1991; Finley, Basri \& Bowyer 1990; Bohlin,
 Colina \& Finley 1995; Kruk et al. 1999; Sing, Holberg \& Dupuis 2002;
 Liebert, Bergeron \& Holberg 2005; Dixon et al. 2007). 
 Comparison with trigonometric parallaxes available for the brightest
 DA white dwarfs gives empirical support to this procedure (Holberg, Bergeron 
 \& Gianninas 2008).

\begin{figure*}
\includegraphics[angle=0,width=8cm]{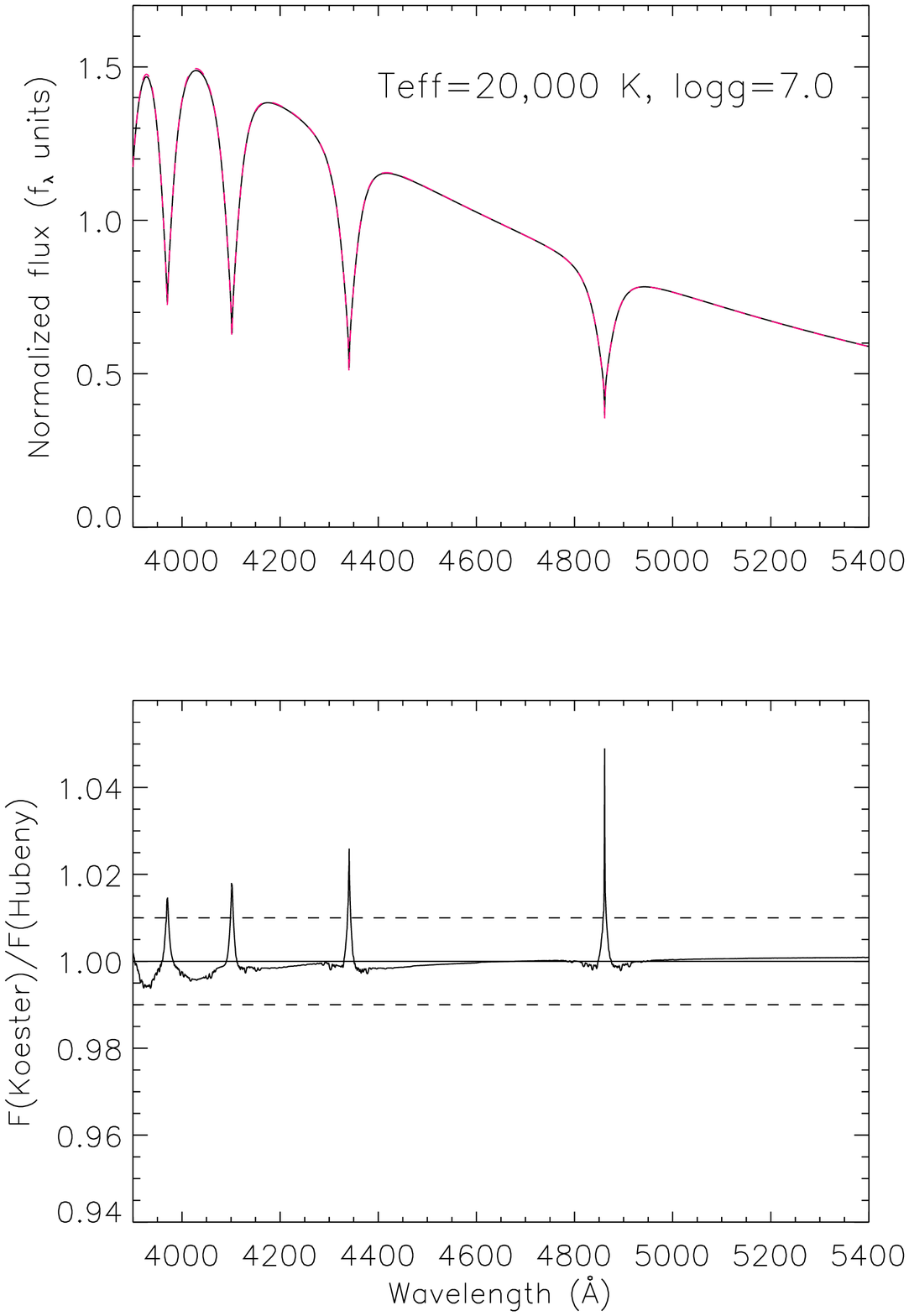}
\includegraphics[angle=0,width=8cm]{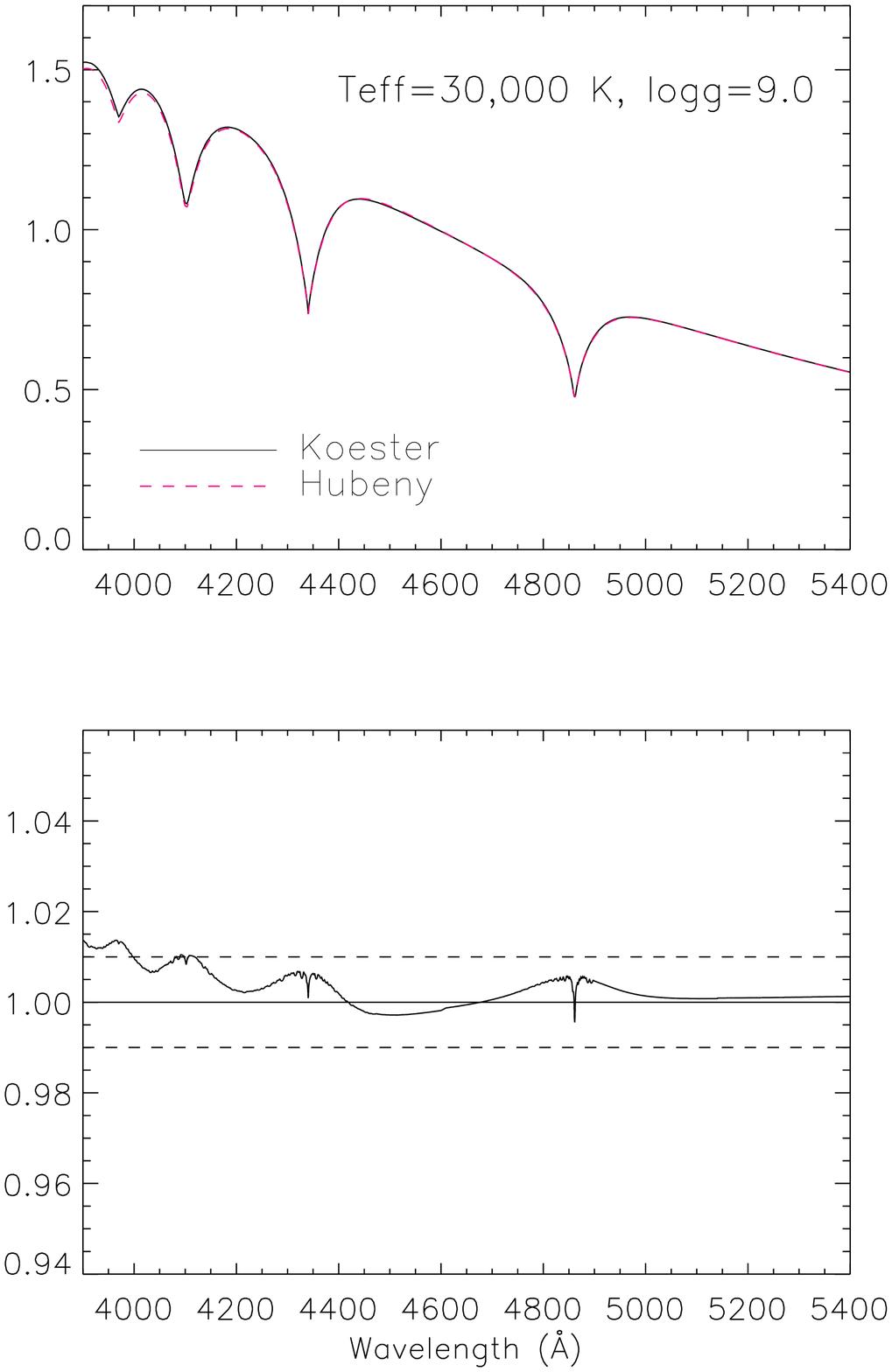}
\caption{A comparison between the model fluxes for DA white dwarfs predicted from the LTE model 
atmospheres by Koester (private communication, black line), 
and Hubeny (red line). Only  the shape of the spectrum is of interest 
for our purposes, and the fluxes are arbitrarily normalized by a constant to
match perfectly at $\log \lambda= 3.67$ (4677.352 \AA). The lower panels show the flux ratios}
\label{f1}
\end{figure*}

Fitting intermediate-resolution spectra of these stars provides estimates 
of the two parameters that define their atmospheric models, effective temperature 
and surface gravity, with a precision better than $\sim$ 1\% and 0.1 dex, respectively.
These uncertainties influence the predicted spectral energy distribution 
to a level of $< 1$ \% 
throughout the optical and near-infrared. 
The solidity of the theoretically predicted 
relative fluxes can be appraised by comparing calculations with  	
independent codes. For an effective temperature of 20,000 K, LTE models computed by
Koester (e.g., Kilic et al. 2007) predict continuum spectra that deviate 
from Tlusty's by only 0.5 \% in the optical and near-infrared, and the 
differences reach only 1\% at 30,000 K, due in part to departures from Local
Thermodynamical Equilibrium (LTE), as illustrates Fig. \ref{f1}.

DA white dwarfs are ideal calibration sources for other reasons.
They are intrinsically faint, and therefore located at small
distances, which renders their observed fluxes unaffected by interstellar
extinction. Comparatively, B- and A-type stars, also with relatively smooth 
continua dominated in the optical and near-infrared by bound-free hydrogen
opacity, present several disadvantages. They show weak metal lines, are
located at larger distances from Earth, and some are surrounded by
disks (Laureijs et al. 2002), which might disturb their spectral shapes 
in the infrared. 
Furthermore, many early type stars rotate fast (see the example of Vega; 
Adelman \& Gulliver 1990) and, although this is hard to detect in the 
spectrum when the rotational axis is aligned with 
the line of sight, rotation can distort significantly their spectral energy 
distribution and broad-band colors (P\'erez Hern\'andez et al. 1999, 
Aufdenberg et al. 2006).

F-type subdwarfs such as the SDSS standard (BD $+17$ 4718; Bohlin \& Gilliland 2004; 
Ram\'{\i}rez et al. 2006) are abundant in the turn-off of the Galactic halo and 
thick-disk populations, but again, they are located at considerably larger distances, 
and their continua are influenced by both hydrogen and H$^{-}$ 	
photoionisation, with the contribution of the latter varying depending on the 
number of free electrons available, which makes their spectral shape 
dependent on their exact metal abundance, increasing the complexity of the model
atmosphere calculations.


In addition to the three fundamental DA stars that set the CALSPEC absolute
fluxes, Hubble Space Telescope (HST) spectrophotometry 
links observations of other stars (some of which are DA) to the  fundamental 
standards, expanding the calibration network to about two dozen objects throughout the sky.
This is important, as three standards are not enough for accurate calibrations, in
particular for ground-based observations where small angular
separations between target and calibrator are a must. Nonetheless, 
this two-step process may result in a loss of accuracy and, due to 
the miscellaneous nature of the secondary calibrators and the resulting large
range in distance and extinction, significant variations 
in quality across the sky.

The link between the secondary
HST flux standards and the three primary ones depends on the calibration of
the time-dependent response of the  HST spectrographs.
The risk of using secondary standards is nowhere more evident than in the recent update
of Vega's HST spectrophotometry. After considering charge
transfer efficiency corrections for the STIS CCD (see Goudfrooij et al. 2006),
the slope in the Paschen continuum was modified by about 2\%. Such a change
implies a decrease in the inferred effective temperature from 9550 K to
9400 K (Bohlin 2007), and in fact, it solves a long-standing discrepancy 
between the HST and the IUE-INES flux scales (Garc\'{\i}a-Gil et al. 2005).

Instead of using secondary standards, based themselves of observations that
are ultimately tied to the three white
dwarfs chosen by Bohlin and collaborators, the flux scale can be set in
a single step using a larger number of well-behaved pure-hydrogen 
hot white dwarfs. But those must be first identified.

While bright flux standards have been favoured in the past,
there is a growing need for fainter standards. The SDSS camera 
and spectrographs cannot observe standards brighter than about 14 magnitude
in their normal mode of operation: transfer delay integration CCD operation 
would not be feasible, and very short exposure times would lead to 
large errors in the fluxes. This situation is likely to become more exhacerbated
for upcoming instrumentation.

The Sloan Digital Sky survey is a major source of  large numbers of 
DA white dwarfs in the range $14<V<21$ (Harris et al. 2003, Kleinman et al. 2004, 
Eisenstein et al. 2006). In this work, we use spectra and photometry in the sixth data 
release (Adelman-McCarthy et al. 2008)
to identify hot DA stars, deriving a homogeneous set of atmospheric parameters and
absolute fluxes for them
to propose an expanded network of DA primary standards. 
Section 2 elaborates on our analysis procedure. Section 3 describes our sample and the 
results. Section 4 describes the determination of the zero points of the absolute flux
scale, and in \S 5 we present our standard candidates. Section 6 revisits the 
determination of atmospheric parameters for the HST primary standards, and \S 7 
closes the paper with a brief summary.

\begin{figure*}
\resizebox{\hsize}{!}{\includegraphics[angle=90]{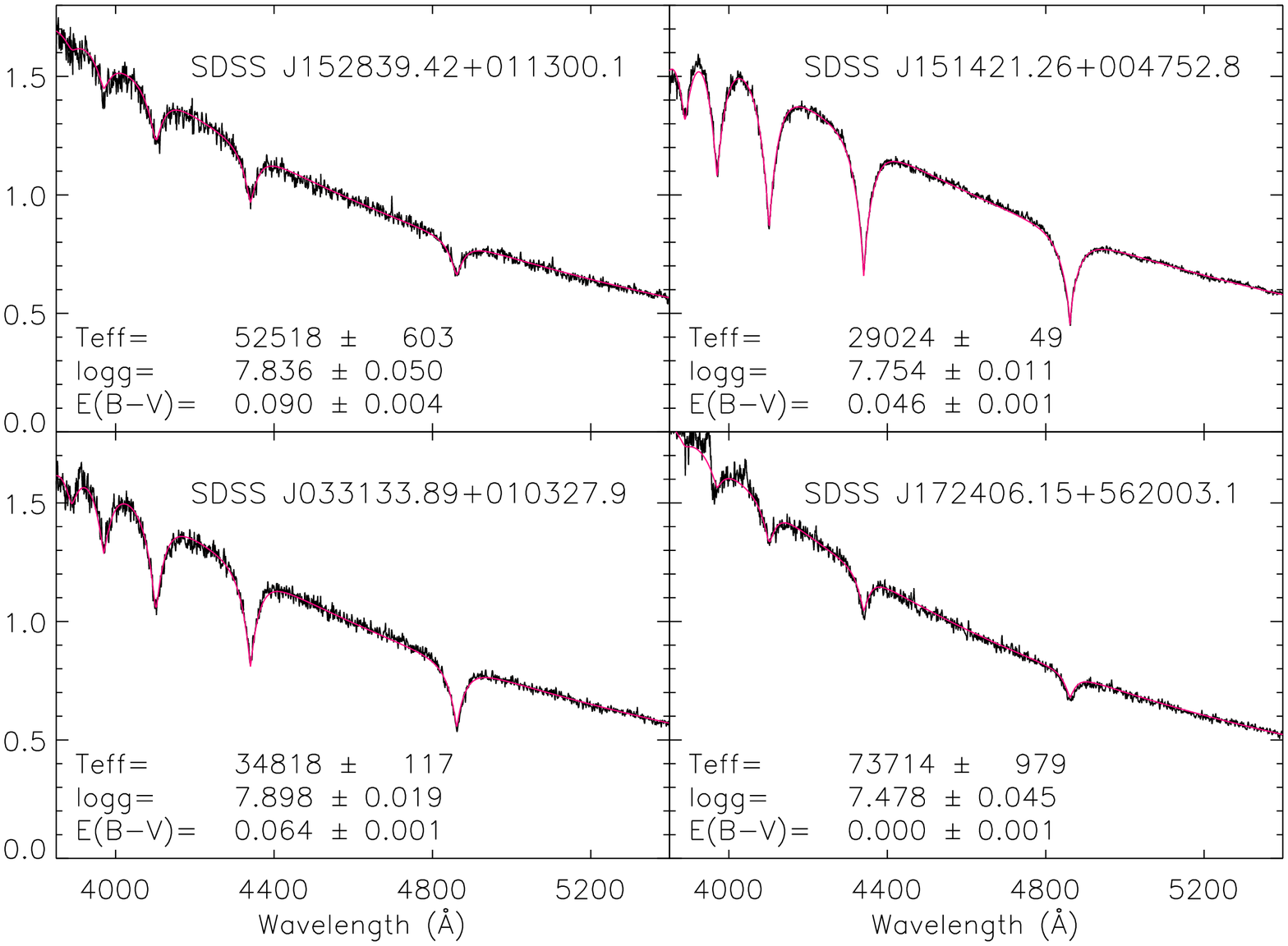}}
\caption{Model fittings to four of the SDSS sample stars. The fluxes are in $f_{\lambda}$ units, 
normalized to have a median of value of one in the range 3850--5400 \AA. Note that the parameters
here do not match those in Table \ref{t1}, as the table gives the average results 
for two types of analyses: those using spectra that preserve continuum information
such as those shown here, and those using continuum-corrected spectra.}
\label{f2}
\end{figure*}

\begin{table*}
\label{t1}
 \centering
\begin{minipage}{140mm}
  \caption{Parameters derived from SDSS spectra.}
  \begin{tabular}{@{}lllll@{}}
  \hline
   Name     &  SDSS  spectrum    &  $T_{\rm eff}$   & $\log g$ & E(B-V)  \\
            &                    & (K)     & (dex)   & (mag) \\
 \hline
SDSS J033133.89+010327.9	 &  spSpec-51810-0415-322  &  35117 ( 695)  &  7.885 (0.031)  &  0.064 (0.031) \\ 
SDSS J033133.89+010327.9	 &  spSpec-51879-0415-337  &  35921 ( 94)  &  7.805 (0.013)  &  0.067 (0.028) \\ 
SDSS J033133.89+010327.9	 &  spSpec-52672-0810-389  &  35358 ( 635)  &  7.811 (0.054)  &  0.084 (0.011) \\ 
SDSS J075058.46+491707.3	 &  spSpec-53089-1779-342  &  24296 ( 1445)  &  7.843 (0.058)  &  0.040 (0.055) \\ 
SDSS J075106.52+301726.4	 &  spSpec-52663-0889-621  &  32973 ( 308)  &  7.812 (0.033)  &  0.000 (0.001) \\ 
SDSS J081126.68+053911.9	 &  spSpec-52934-1295-168  &  27912 ( 557)  &  7.808 (0.067)  &  0.073 (0.073) \\ 
SDSS J081234.94+040852.1	 &  spSpec-52641-1184-347  &  27420 ( 399)  &  7.753 (0.044)  &  0.031 (0.014) \\ 
SDSS J082346.15+245345.7	 &  spSpec-52962-1585-605  &  33892 ( 212)  &  7.709 (0.042)  &  0.011 (0.011) \\ 
SDSS J084537.74+065346.2	 &  spSpec-52964-1298-151  &  22062 ( 663)  &  8.410 (0.038)  &  0.040 (0.055) \\ 
SDSS J091558.19+201606.2	 &  spSpec-53699-2288-223  &  24409 ( 475)  &  8.230 (0.064)  &  0.014 (0.014) \\ 
SDSS J091601.87+200758.1	 &  spSpec-53699-2288-194  &  32102 ( 401)  &  7.184 (0.020)  &  0.027 (0.068) \\ 
SDSS J092010.55+045721.1	 &  spSpec-52707-0991-221  &  62729 ( 2648)  &  7.266 (0.097)  &  0.051 (0.044) \\ 
SDSS J094203.19+544630.2	 &  spSpec-51991-0556-112  &  28746 ( 388)  &  7.877 (0.019)  &  0.000 (0.002) \\ 
SDSS J094940.37+032425.5	 &  spSpec-52286-0571-257  &  49492 ( 2205)  &  7.432 (0.129)  &  0.032 (0.048) \\ 
SDSS J095230.44+114202.3	 &  spSpec-53054-1743-184  &  27294 ( 330)  &  7.855 (0.018)  &  0.004 (0.004) \\ 
SDSS J095245.59+020938.9	 &  spSpec-51908-0481-506  &  44492 ( 1453)  &  7.639 (0.047)  &  0.000 (0.095) \\ 
SDSS J100222.50+292755.0	 &  spSpec-53436-1950-124  &  70874 ( 5659)  &  7.372 (0.200)  &  0.000 (0.095) \\ 
SDSS J100543.92+304744.7	 &  spSpec-53358-1953-378  &  66984 ( 4732)  &  7.511 (0.036)  &  0.001 (0.094) \\ 
SDSS J100614.76+441906.3	 &  spSpec-52703-0942-636  &  55721 ( 3152)  &  7.639 (0.147)  &  0.000 (0.095) \\ 
SDSS J101328.17+061207.4	 &  spSpec-52641-0996-066  &  49416 ( 2060)  &  7.740 (0.026)  &  0.004 (0.091) \\ 
SDSS J103743.48+485720.8	 &  spSpec-52354-0875-133  &  21179 ( 213)  &  8.130 (0.017)  &  0.054 (0.041) \\ 
SDSS J103907.38+081840.9	 &  spSpec-52734-1240-097  &  23564 ( 402)  &  7.386 (0.069)  &  0.050 (0.045) \\ 
SDSS J104332.62+445329.0	 &  spSpec-52992-1431-591  &  50379 ( 1496)  &  7.618 (0.049)  &  0.001 (0.094) \\ 
SDSS J104419.01+405553.0	 &  spSpec-53035-1433-405  &  27852 ( 229)  &  7.718 (0.074)  &  0.003 (0.086) \\ 
SDSS J110634.39+073712.2	 &  spSpec-52723-1004-388  &  36110 ( 1728)  &  7.703 (0.093)  &  0.082 (0.013) \\ 
SDSS J114152.63+253533.9	 &  spSpec-53856-2505-581  &  45697 ( 3797)  &  7.556 (0.103)  &  0.019 (0.076) \\ 
SDSS J120525.01+303444.7	 &  spSpec-53729-2225-617  &  29958 ( 227)  &  7.766 (0.046)  &  0.000 (0.013) \\ 
SDSS J121205.11+140801.8	 &  spSpec-53466-1765-136  &  21154 ( 316)  &  8.306 (0.033)  &  0.019 (0.019) \\ 
SDSS J121845.69+264831.8	 &  spSpec-53816-2231-133  &  26341 ( 384)  &  7.790 (0.010)  &  0.020 (0.045) \\ 
SDSS J122336.20+412242.7	 &  spSpec-53112-1452-206  &  23236 ( 700)  &  7.937 (0.034)  &  0.000 (0.003) \\ 
SDSS J124407.67+582351.9	 &  spSpec-52765-1317-405  &  32179 ( 143)  &  7.896 (0.016)  &  0.000 (0.095) \\ 
SDSS J125217.15+154443.2	 &  spSpec-53171-1770-567  &  26290 ( 611)  &  7.239 (0.055)  &  0.014 (0.016) \\ 
SDSS J130234.44+101239.0	 &  spSpec-53883-1793-508  &  43656 ( 319)  &  7.760 (0.071)  &  0.020 (0.073) \\ 
SDSS J132232.12+641545.8	 &  spSpec-52056-0603-477  &  29074 ( 89)  &  7.339 (0.100)  &  0.000 (0.002) \\ 
SDSS J132434.39+072525.3	 &  spSpec-53556-1799-497  &  27648 ( 83)  &  7.851 (0.079)  &  0.005 (0.005) \\ 
SDSS J133207.33+665453.4	 &  spSpec-51989-0497-346  &  27701 ( 1373)  &  7.862 (0.061)  &  0.000 (0.002) \\ 
SDSS J133514.52+505012.3	 &  spSpec-53433-1669-350  &  38139 ( 2010)  &  7.765 (0.173)  &  0.048 (0.047) \\ 
SDSS J134430.11+032423.2	 &  spSpec-52025-0529-572  &  26187 ( 101)  &  7.825 (0.010)  &  0.028 (0.028) \\ 
SDSS J140327.76+002119.6	 &  spSpec-51942-0301-626  &  65315 ( 4775)  &  7.461 (0.198)  &  0.019 (0.076) \\ 
SDSS J142020.80+521549.3	 &  spSpec-52725-1045-609  &  24548 ( 480)  &  7.757 (0.022)  &  0.055 (0.055) \\ 
SDSS J143059.88+100142.9	 &  spSpec-53533-1709-187  &  24910 ( 576)  &  7.824 (0.046)  &  0.000 (0.032) \\ 
SDSS J143105.74+042215.6	 &  spSpec-52027-0585-495  &  23740 ( 565)  &  7.755 (0.093)  &  0.025 (0.025) \\ 
SDSS J143315.92+252853.1	 &  spSpec-53827-2135-156  &  23860 ( 290)  &  7.137 (0.012)  &  0.010 (0.010) \\ 
SDSS J143443.25+533521.2	 &  spSpec-52764-1326-639  &  22667 ( 188)  &  7.755 (0.044)  &  0.033 (0.033) \\ 
SDSS J144814.08+282511.7	 &  spSpec-53764-2141-127  &  24478 ( 1657)  &  8.332 (0.047)  &  0.000 (0.001) \\ 
SDSS J145415.84+551152.3	 &  spSpec-52353-0792-285  &  28680 ( 332)  &  8.285 (0.025)  &  0.000 (0.002) \\ 
SDSS J145600.81+574150.8	 &  spSpec-52056-0610-288  &  31989 ( 341)  &  7.634 (0.050)  &  0.000 (0.002) \\ 
SDSS J150045.01+621107.2	 &  spSpec-52339-0609-039  &  23452 ( 481)  &  7.765 (0.013)  &  0.000 (0.003) \\ 
SDSS J150422.29+621718.6	 &  spSpec-52055-0611-538  &  63549 ( 2154)  &  7.548 (0.221)  &  0.000 (0.095) \\ 
SDSS J151421.26+004752.8	 &  spSpec-51689-0312-371  &  29080 ( 121)  &  7.749 (0.013)  &  0.046 (0.008) \\ 
SDSS J152041.96+495140.8	 &  spSpec-52751-1166-270  &  29809 ( 67)  &  7.333 (0.106)  &  0.000 (0.095) \\ 
SDSS J152839.42+011300.1	 &  spSpec-51641-0314-331  &  54435 ( 4570)  &  7.730 (0.184)  &  0.090 (0.006) \\ 
SDSS J160839.07+074542.5	 &  spSpec-53498-1730-199  &  23508 ( 89)  &  8.464 (0.060)  &  0.026 (0.026) \\ 
SDSS J165318.76+371027.2	 &  spSpec-52433-0820-328  &  20599 ( 1109)  &  8.198 (0.010)  &  0.000 (0.002) \\ 
SDSS J165851.11+341853.3	 &  spSpec-52435-0972-475  &  58350 ( 866)  &  7.635 (0.023)  &  0.000 (0.095) \\ 
SDSS J170331.62+223251.3	 &  spSpec-53462-1688-188  &  25687 ( 1629)  &  7.840 (0.018)  &  0.076 (0.076) \\ 
SDSS J172406.14+562003.1	 &  spSpec-51818-0358-318  &  36442 ( 103)  &  7.267 (0.016)  &  0.000 (0.095) \\ 
SDSS J212412.14+110415.7	 &  spSpec-52466-0730-392  &  22759 ( 237)  &  7.772 (0.078)  &  0.011 (0.011) \\ 
SDSS J214001.05-075052.2	 &  spSpec-52824-1177-480  &  31605 ( 402)  &  7.753 (0.059)  &  0.027 (0.068) \\ 
\hline
\end{tabular}
\end{minipage}
\end{table*}

\section{Analysis}
\label{analysis}

We focus on DA stars with a surface temperature warmer than 20,000 K. This avoids
known problems with models for DA cooler than about 12,000 K 
for which unexpectedly large surface gravities are derived from the 
analysis of Balmer lines
(see, e.g. Bergeron, Saffer \& Liebert 1992, Eisenstein et al. 2006). 
Our choice of temperatures  avoids the presence 
of  convective energy transport in these atmospheres, allowing us to
safely assume radiative balance. 

We use an unpublished grid of model fluxes for DA stars previously 
calculated by Hubeny. The model atmospheres  contain only hydrogen and 
are computed, in Non-LTE,  with the code Tlusty (Hubeny \& Lanz 1995, 
Lanz \& Hubeny 1995). The 
code, and the hydrogen model atom used are publicly available from the Tlusty web 
site\footnote{\tt http://nova.astro.umd.edu}.
We considered surface gravities in the range $7 < \log g < 9.5$ (c.g.s. units). 
Among the most important ingredients, H line profiles were calculated with the Stark 
broadening tables of Lemke (1997), and the level dissolution, occupation probabilities, 
and the pseudo-continuum opacities were taken from Hubeny, Hummer \& Lanz (1994).

Our spectral analysis uses the same code and principles outlined  
in Kilic et al. (2007), but we have added one more dimension to the 
problem in order to account for reddening. For each star we determine the  trio  
$(T_{\rm eff}, \log g, E(B-V))$ of a model that matches best the observations 
in the region $385-540$ nm (air wavelengths). The spectra are either 
normalized by a constant, taken as the median value of the flux in the 
selected spectral window, or by a polynomial model of the star's pseudocontinuum
(see \S \ref{sample} for more details).
A wavelength grid equidistant
in $\log \lambda$ was employed. The theoretical fluxes are computed 
for a discrete grid with steps of 2,000 K and 0.5 dex for $T_{\rm eff}$ and $\log g$, 
respectively, and smoothed to a FWHM ($\equiv \delta\lambda$) 
resolving power of $\lambda/\delta\lambda = 2000$, 
which approximately corresponds to that of the SDSS spectra. Reddening is
introduced with a discrete grid sampling $0<E(B-V)<0.095$ in steps of 0.005 mag,
using the prescription of Fitzpatrick (1999) with  
$R\equiv A_V/E(B-V) = 3.1$.

The optimization is based on the Nelder-Mead method (Nelder \& Mead 1965)
assisted by Bezier quadratic interpolation in the grid of synthetic fluxes
(see, e.g., Auer 2003). Internal uncertainties are estimated, assuming 
the errors in the fluxes are normally distributed, from the diagonal 
of the inverse of the curvature matrix.

\section{Sample Selection}
\label{sample}

With the goal of identifying solid DA candidates for a network of flux
standards, we scanned the SDSS DR6 photometric catalog in search for hot objects
satisfying $(u-g)_0<0.6$ and $(g-r)_0<-0.2$. We also imposed the constraint of
a moderate estimated extinction towards the sources: $A_g < 0.827$,
with $A_g$ from the dust maps of Schlegel, Finkbeiner \& Davis (1998). 
This is approximately
equivalent to $E(B-V)< 0.27$ mag, but this value would only apply to objects outside
the Milky Way, or at least beyond the extinction layer; 
we will later restrict our analysis to stars with $E(B-V)<0.1$. 
To allow a reliable
determination of the stellar parameters, necessary to assign unique 
theoretical spectral energy distributions, we also limited our search to
sources in the range $14<g<17$. Our query to the SDSS 
Catalog Archive Server (CAS; Thakar et al. 2008) 
returned 598 objects\footnote{The query returned,
in fact,  599 sources, but one file ("spSpec-53388-1937-316.fit") had an anomalous format.}.

Viable DA candidates must show spectra that are well matched by DA models.
Thus, we downloaded the spectra for all the selected targets and analyzed 
them following the prescription in \S \ref{analysis}.
The quality of the fittings was assessed from the $\chi^2$ statistics, 
and used to select the sources that were
closely approximated by theoretical fluxes for DA stars. Although white
dwarfs are very dim, at the faint magnitudes covered by the SDSS hot
DA are at distances of a few hundred pc, and extinction becomes significant,
at least when aiming at modeling flux distributions with high accuracy.

\begin{figure*}
\resizebox{\hsize}{!}{\includegraphics[angle=90]{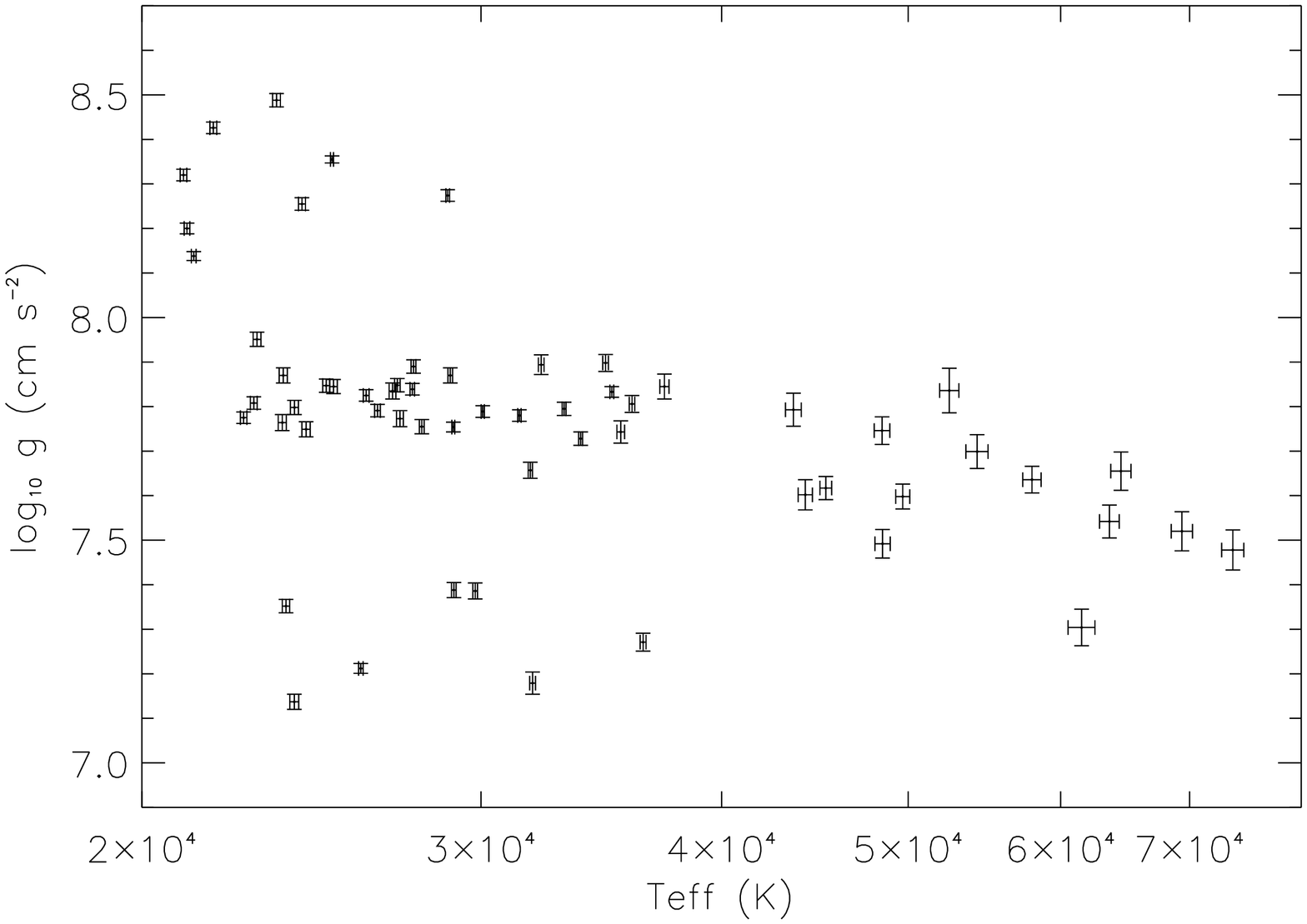}}
\caption{$T\_{\rm eff}$ and $\log g$ derived from the analysis of spectra with 
continuum information such as those illustrated in Fig. \ref{f2}. The uncertainties
are internal and do not consider systematic errors in the models or the observations.
 Note that a non-linear scale is being used for the abscissae for the sake of clarity.}
\label{f3}
\end{figure*}

The radial velocities determined by the SDSS  pipeline were used to 
correct the Doppler shifts in the spectra before the analysis, and
both the model fluxes and the observed spectra were normalized to 
have a median flux equal to one in the wavelength range used in the analysis.
Only 57 objects which spectra  
were matched by the model fluxes with $\chi^2 < 0.63$ and with derived
atmospheric parameters $21,000 < T_{\rm eff} <85,000 $K and $7.1<\log g< 9.4$, 
and limited extinction, $0.0 \le E(B-V) < 0.095$, 
were retained for further analysis. 
Fig. \ref{f2} shows a few representative examples of the fittings
in the case where the spectra are normalized by their median fluxes
in the analysis window. As mentioned earlier, we also analyzed the
spectra after rectifying the continuum shape.

In Fig. \ref{f3} we show the determined atmospheric parameters for the sample.
As expected, the derived gravities decrease smoothly for hotter stars (see,
e.g. Althaus et al. 2005). At least in the range $20,000 < T_{\rm eff}<40,000$ K,
three clear groups are apparent. These likely correspond to the three mass 
clusters found for DA white dwarfs in studies using the Palomar Green
survey (Liebert et al. 2005), 
and in earlier SDSS data releases (e.g. 
Kepler et al. 2007, DeGennaro et al. 2008). Perhaps due to our strict
selection criteria for the quality of the fittings, and the resulting
minute error bars, the three groups appear here more cleanly separated
than in previous reports. 
We note that heavy elements are  more commonly
found in the atmospheres of  DA stars hotter 
than 50,000 K (Barstow et al. 2003b), 
making such stars statistically 
more vulnerable to underestimated effective temperatures when their  hydrogen
lines are interpreted with pure-H model atmospheres 
(Barstow, Hubeny \& Holberg 1998). The group of low-mass white dwarfs 
with $T_{\rm eff}< 40,000$ K and $\log g<7.5$ are probably the result of 
stellar evolution in close binary systems, which makes their radial velocities 
likely to vary.

Kilic et al. (2007) considered both continuum-corrected fluxes and relative fluxes 
on the determination of
atmospheric parameters for DA white dwarfs from medium resolution MMT spectra. 
Their conclusion was that using
 well-calibrated relative fluxes helped considerably to reduce degeneracies,
and no significant systematic differences in the inferred surface gravities were apparent.
 The continuum-corrected 
spectra are immune to the distortions of interstellar extinction, and therefore
similar surface temperatures obtained by the two methods places a limit to
the interstellar absorption that the spectra in our sample have suffered.

\begin{figure*}
\includegraphics[angle=0,width=8cm]{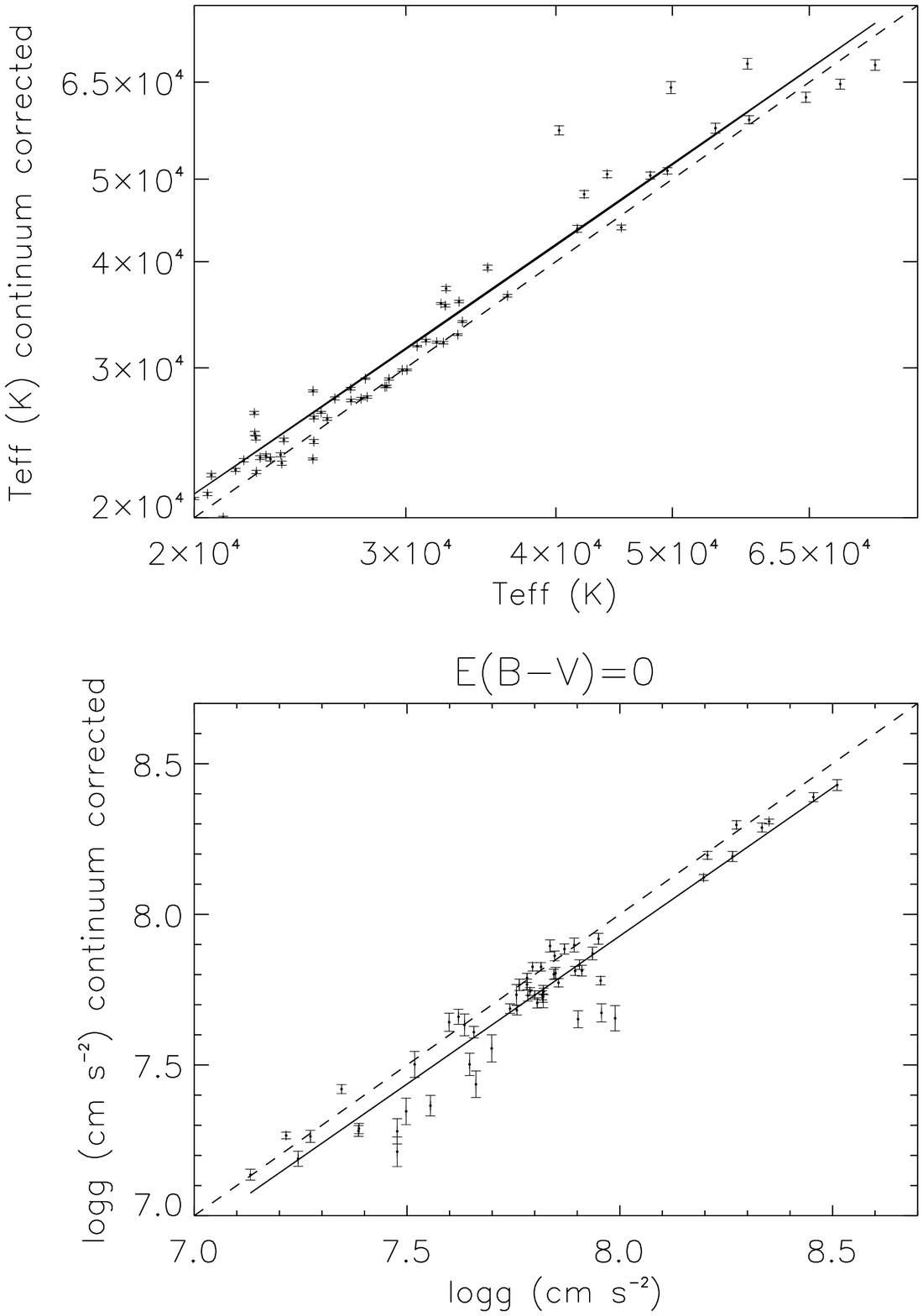}
\includegraphics[angle=0,width=8cm]{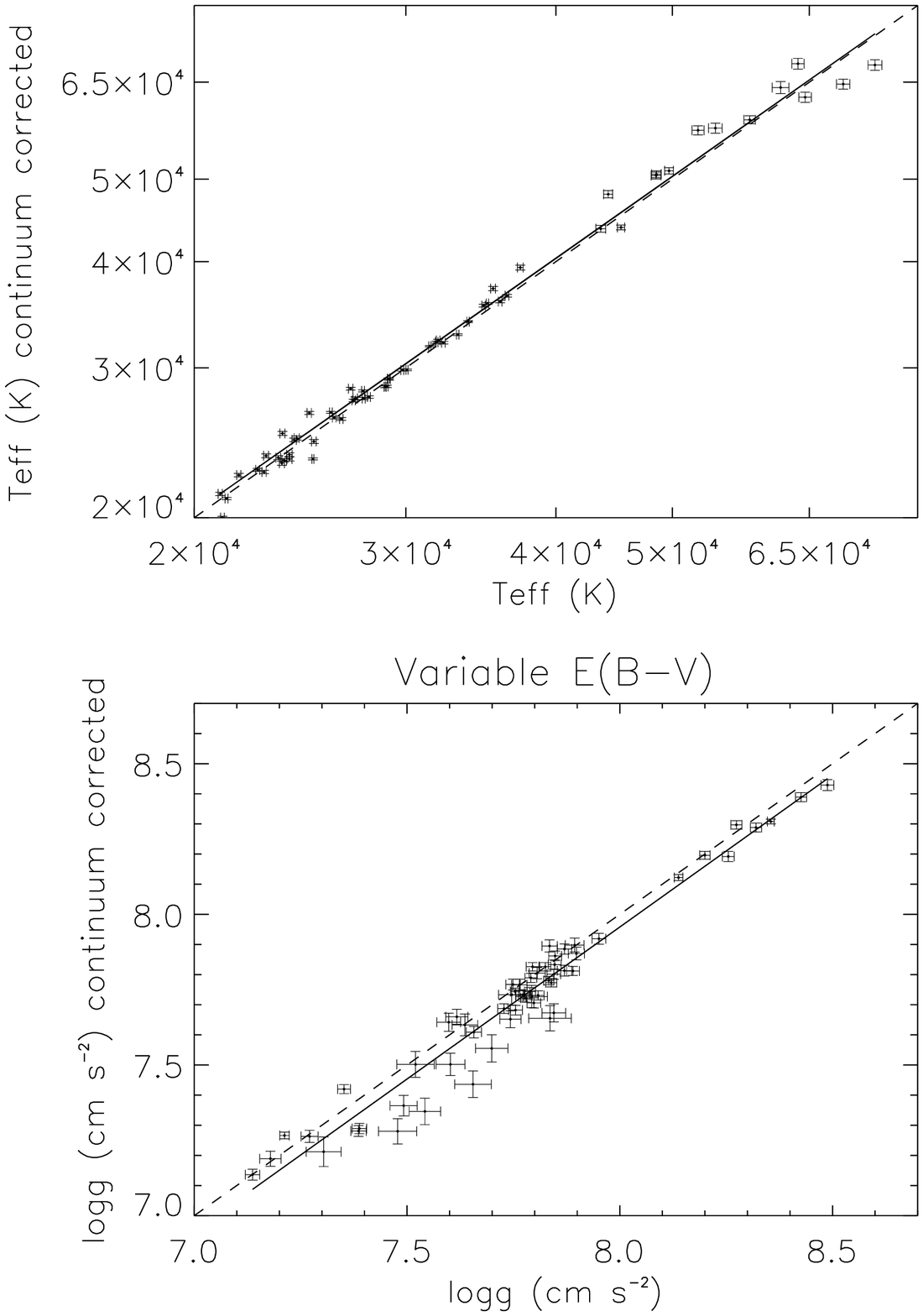}
\caption{Comparison between the atmospheric parameters  retrieved for our SDSS sample 
stars using  continuum-corrected spectra and spectra that preserve the continuum shape. 
The left-hand panels correspond to the case when reddening
is neglected, while the right-hand panels correspond to an analysis that accounts
for reddening}
\label{f4}
\end{figure*}

We repeated such an experiment for the particular case of the SDSS spectra considered
in this work. Reddening needs not be considered when analyzing the continuum-corrected
spectra. However, 
if reddening is neglected (forcing $E(B-V)=0$) in the analysis of 
spectra containing continuum information, a 
systematic trend is apparent. This is illustrated in Fig. \ref{f4}, where the derived 
$T_{\rm eff}$ from continuum-corrected spectra are typically higher,
in particular for the hotter stars, 
in line with the expected signature of interstellar
reddening. When reddening is considered, such trend is effectively removed, 
and the agreement between the two methods greatly improved, 
as shown in Fig. \ref{f4}. Systematic differences are also present in $\log g$, 
for both cases (with and without reddening corrections), but
these are also reduced when reddening is accounted for and 
have nevertheless a far smaller impact on the absolute fluxes.

The results based on spectra with continuum information are more precise, 
but are also more exposed to systematic errors in the SDSS flux calibration.
Our analysis of such spectra, when reddening is considered, returns parameters that
are on the same scale as those from continuum-rectified spectra.
Therefore, we adopt as the final parameters ($T_{\rm eff}$ and $\log g$) 
the weighted average of the two analyses.
It is also noticeable in Fig. \ref{f4} that the internal random errors derived
from each method are not always consistent. Conservatively, we adopt as
$1-\sigma$ uncertainties the sum in quadrature of the estimated 
random uncertainties and the systematic difference between the
two methods. By fitting the spectra with continuum information with
the $T_{\rm eff}$ and $\log g$ fixed to the average values and allowing 
$E(B-V)$ to vary, we derive an estimate for the uncertainty in $E(B-V)$ 
for each star, which is again added in quadrature with the internal 
uncertainty to obtain the final estimate. 
The final adopted parameters are given in Table \ref{t1}. 
Typical (median values) 
uncertainties are 1.8 \%, 0.047 dex, and 0.032 mag in $T_{\rm eff}$, $\log g$,
and $E(B-V)$, respectively. 
These uncertainties in $T_{\rm eff}$ and $\log g$ 
are slightly larger than those inferred by Liebert et al. (2005) for their analysis
of spectra with typically higher signal-to-noise ratios but lower resolution.
There are three spectra of the star SDSS J033133.89$+$010327.9, and 
the derived parameters are fairly consistent.

Most of the stars in our sample (exactly 43; see Table \ref{t1}), are 
included in the SDSS-DR4 catalog of DA stars of Eisenstein et al. (2006). 
Between 20,000 and 30,000 K, the $T_{\rm eff}$ values of Eisenstein et al.
are lower than ours by a few percent, and the trend reverses for
warmer temperatures, with the Eisenstein et al. values being up to
$\sim 10$ \% higher than ours at about 65,000 K. 
We  find good agreement between our
surface gravities and those from the SDSS-DR4 catalog at the
lowest temperatures in our sample ($\sim 20,000$ K), but growing
discrepancies at warmer temperatures, with the SDSS-DR4 values
being $\sim$ 0.25 dex higher around 65,000 K. There are 15 stars
included in Table \ref{t1} which 
are also part in the catalog of DA stars from
the Palomar Green survey (Liebert et al. 2005). The Liebert et al.
surface temperatures are also a few percent lower than ours in the 
range $20,000 <T_{\rm eff}< 30,000$ K, but they are lower by 
as much as 10 \% at about 60,000 K. Their surface gravities 
are systematically  higher by $0.08 \pm 0.02$ dex, and it should be
noted that a similar discrepancy was also found by Liebert et al.
when they compared their gravities 
with several other investigations (Finley et al. 1997,
Marsh et al. 1997, Homeier et al. 1998, Koester et al. 2001). 
In the hot end, our temperature scale is intermediate between 
those of Eisenstein et al. and Liebert et al., which differ from each other by
$\sim 20$ \%.

\begin{table*}
 \centering
 \begin{minipage}{140mm}
  \caption{SDSS Photometry and zero points.}
  \begin{tabular}{@{}llllllll@{}}
  \hline
   Name     &  $u$ & $g$ & $r$ & $\delta u$ & $\delta g$ & $\delta r$ & $\delta$ ($g$ and $r$)  \\
 \hline
SDSS J033133.89+010327.9  &  16.183 (0.022)  &  16.432 (0.013)  &  16.876 (0.012)  &  56.633  &  56.587  &  56.632  &  56.610 (0.023)  \\ 
SDSS J033133.89+010327.9  &  16.183 (0.022)  &  16.432 (0.013)  &  16.876 (0.012)  &  56.717  &  56.651  &  56.675  &  56.663 (0.012)  \\ 
SDSS J033133.89+010327.9  &  16.183 (0.022)  &  16.432 (0.013)  &  16.876 (0.012)  &  56.747  &  56.678  &  56.701  &  56.689 (0.012)  \\ 
SDSS J075058.46+491707.3  &  16.664 (0.015)  &  16.740 (0.018)  &  17.117 (0.018)  &  56.294  &  56.263  &  56.267  &  56.265 (0.002)  \\ 
SDSS J075106.52+301726.4  &  15.649 (0.015)  &  15.917 (0.020)  &  16.310 (0.022)  &  56.351  &  56.265  &  56.169  &  56.217 (0.048)  \\ 
SDSS J081126.68+053911.9  &  16.261 (0.016)  &  16.416 (0.012)  &  16.778 (0.010)  &  56.126  &  56.110  &  56.103  &  56.107 (0.004)  \\ 
SDSS J081234.94+040852.1  &  16.723 (0.016)  &  16.860 (0.012)  &  17.302 (0.009)  &  56.737  &  56.674  &  56.700  &  56.687 (0.013)  \\ 
SDSS J082346.15+245345.7  &  15.273 (0.028)  &  15.552 (0.025)  &  16.039 (0.014)  &  55.984  &  55.911  &  55.916  &  55.914 (0.003)  \\ 
SDSS J084537.74+065346.2  &  16.539 (0.018)  &  16.637 (0.016)  &  16.972 (0.014)  &  55.932  &  55.967  &  55.968  &  55.968 (0.001)  \\ 
SDSS J091558.19+201606.2  &  16.529 (0.013)  &  16.590 (0.015)  &  17.026 (0.010)  &  56.316  &  56.221  &  56.261  &  56.241 (0.020)  \\ 
SDSS J091601.87+200758.1  &  16.674 (0.013)  &  16.941 (0.015)  &  17.395 (0.010)  &  57.160  &  57.130  &  57.123  &  57.126 (0.003)  \\ 
SDSS J092010.55+045721.1  &  16.274 (0.019)  &  16.673 (0.016)  &  17.188 (0.010)  &  57.652  &  57.624  &  57.659  &  57.642 (0.017)  \\ 
SDSS J094203.19+544630.2  &  16.714 (0.013)  &  16.934 (0.017)  &  17.398 (0.016)  &  57.027  &  56.975  &  56.977  &  56.976 (0.001)  \\ 
SDSS J094940.37+032425.5  &  16.454 (0.020)  &  16.787 (0.016)  &  17.301 (0.014)  &  57.656  &  57.571  &  57.595  &  57.583 (0.012)  \\ 
SDSS J095230.44+114202.3  &  16.317 (0.022)  &  16.472 (0.016)  &  16.869 (0.019)  &  56.454  &  56.379  &  56.331  &  56.355 (0.024)  \\ 
SDSS J095245.59+020938.9  &  16.026 (0.015)  &  16.347 (0.021)  &  16.832 (0.018)  &  57.250  &  57.135  &  57.099  &  57.117 (0.018)  \\ 
SDSS J100222.50+292755.0  &  15.677 (0.010)  &  16.136 (0.011)  &  16.669 (0.016)  &  57.428  &  57.395  &  57.382  &  57.388 (0.007)  \\ 
SDSS J100543.92+304744.7  &  15.967 (0.023)  &  16.346 (0.018)  &  16.904 (0.013)  &  57.651  &  57.544  &  57.558  &  57.551 (0.007)  \\ 
SDSS J100614.76+441906.3  &  16.401 (0.024)  &  16.824 (0.014)  &  17.369 (0.014)  &  57.892  &  57.846  &  57.856  &  57.851 (0.005)  \\ 
SDSS J101328.17+061207.4  &  15.845 (0.018)  &  16.280 (0.019)  &  16.754 (0.021)  &  57.180  &  57.165  &  57.116  &  57.141 (0.024)  \\ 
SDSS J103743.48+485720.8  &  15.275 (0.021)  &  15.303 (0.021)  &  15.648 (0.015)  &  54.475  &  54.501  &  54.533  &  54.517 (0.016)  \\ 
SDSS J103907.38+081840.9  &  16.222 (0.007)  &  16.180 (0.017)  &  16.595 (0.021)  &  55.701  &  55.604  &  55.656  &  55.630 (0.026)  \\ 
SDSS J104332.62+445329.0  &  16.271 (0.012)  &  16.673 (0.014)  &  17.183 (0.016)  &  57.644  &  57.591  &  57.573  &  57.582 (0.009)  \\ 
SDSS J104419.01+405553.0  &  16.660 (0.028)  &  16.832 (0.040)  &  17.237 (0.017)  &  56.855  &  56.788  &  56.739  &  56.763 (0.024)  \\ 
SDSS J110634.39+073712.2  &  16.548 (0.011)  &  16.878 (0.030)  &  17.305 (0.025)  &  57.047  &  57.076  &  57.098  &  57.087 (0.011)  \\ 
SDSS J114152.63+253533.9  &  16.620 (0.026)  &  16.990 (0.029)  &  17.480 (0.016)  &  57.787  &  57.736  &  57.725  &  57.731 (0.005)  \\ 
SDSS J120525.01+303444.7  &  15.697 (0.024)  &  15.877 (0.022)  &  16.300 (0.014)  &  56.132  &  56.016  &  55.965  &  55.990 (0.026)  \\ 
SDSS J121205.11+140801.8  &  16.810 (0.012)  &  16.811 (0.020)  &  17.153 (0.014)  &  56.186  &  56.139  &  56.132  &  56.136 (0.003)  \\ 
SDSS J121845.69+264831.8  &  16.320 (0.011)  &  16.469 (0.012)  &  16.861 (0.021)  &  56.272  &  56.237  &  56.211  &  56.224 (0.013)  \\ 
SDSS J122336.20+412242.7  &  16.887 (0.019)  &  16.926 (0.026)  &  17.285 (0.020)  &  56.593  &  56.512  &  56.465  &  56.488 (0.024)  \\ 
SDSS J124407.67+582351.9  &  16.589 (0.017)  &  16.846 (0.026)  &  17.300 (0.018)  &  57.231  &  57.144  &  57.114  &  57.129 (0.015)  \\ 
SDSS J125217.15+154443.2  &  14.339 (0.033)  &  14.403 (0.029)  &  14.800 (0.021)  &  54.294  &  54.189  &  54.150  &  54.169 (0.020)  \\ 
SDSS J130234.44+101239.0  &  16.627 (0.018)  &  16.988 (0.018)  &  17.470 (0.024)  &  57.729  &  57.677  &  57.663  &  57.670 (0.007)  \\ 
SDSS J132232.12+641545.8  &  16.054 (0.017)  &  16.268 (0.024)  &  16.658 (0.016)  &  56.381  &  56.333  &  56.249  &  56.291 (0.042)  \\ 
SDSS J132434.39+072525.3  &  16.417 (0.019)  &  16.581 (0.018)  &  17.006 (0.023)  &  56.586  &  56.513  &  56.491  &  56.502 (0.011)  \\ 
SDSS J133207.33+665453.4  &  16.669 (0.017)  &  16.812 (0.026)  &  17.342 (0.015)  &  56.869  &  56.767  &  56.844  &  56.806 (0.039)  \\ 
SDSS J133514.52+505012.3  &  16.484 (0.017)  &  16.781 (0.013)  &  17.262 (0.013)  &  57.247  &  57.190  &  57.221  &  57.205 (0.016)  \\ 
SDSS J134430.11+032423.2  &  16.482 (0.015)  &  16.603 (0.018)  &  17.005 (0.016)  &  56.380  &  56.328  &  56.323  &  56.326 (0.002)  \\ 
SDSS J140327.76+002119.6  &  16.410 (0.022)  &  16.856 (0.017)  &  17.376 (0.018)  &  57.981  &  57.962  &  57.963  &  57.963 (0.000)  \\ 
SDSS J142020.80+521549.3  &  16.753 (0.024)  &  16.855 (0.018)  &  17.273 (0.015)  &  56.334  &  56.342  &  56.401  &  56.371 (0.030)  \\ 
SDSS J143059.88+100142.9  &  16.537 (0.022)  &  16.629 (0.019)  &  17.044 (0.017)  &  56.428  &  56.355  &  56.343  &  56.349 (0.006)  \\ 
SDSS J143105.74+042215.6  &  16.808 (0.014)  &  16.791 (0.020)  &  17.174 (0.011)  &  56.443  &  56.325  &  56.321  &  56.323 (0.002)  \\ 
SDSS J143315.92+252853.1  &  16.810 (0.030)  &  16.772 (0.021)  &  17.168 (0.015)  &  56.509  &  56.373  &  56.351  &  56.362 (0.011)  \\ 
SDSS J143443.25+533521.2  &  15.921 (0.017)  &  15.942 (0.021)  &  16.248 (0.012)  &  55.390  &  55.354  &  55.294  &  55.324 (0.030)  \\ 
SDSS J144814.08+282511.7  &  14.254 (0.025)  &  14.398 (0.023)  &  14.802 (0.014)  &  54.121  &  54.088  &  54.081  &  54.085 (0.003)  \\ 
SDSS J145415.84+551152.3  &  15.584 (0.013)  &  15.813 (0.014)  &  16.275 (0.016)  &  55.908  &  55.851  &  55.859  &  55.855 (0.004)  \\ 
SDSS J145600.81+574150.8  &  15.951 (0.022)  &  16.191 (0.020)  &  16.679 (0.016)  &  56.569  &  56.475  &  56.476  &  56.476 (0.000)  \\ 
SDSS J150045.01+621107.2  &  16.909 (0.016)  &  16.961 (0.020)  &  17.326 (0.020)  &  56.632  &  56.565  &  56.518  &  56.541 (0.024)  \\ 
SDSS J150422.29+621718.6  &  16.328 (0.017)  &  16.808 (0.018)  &  17.277 (0.018)  &  57.962  &  57.958  &  57.886  &  57.922 (0.036)  \\ 
SDSS J151421.26+004752.8  &  15.483 (0.012)  &  15.687 (0.012)  &  16.103 (0.012)  &  55.603  &  55.579  &  55.583  &  55.581 (0.002)  \\ 
SDSS J152041.96+495140.8  &  16.037 (0.040)  &  16.268 (0.026)  &  16.722 (0.034)  &  56.442  &  56.392  &  56.368  &  56.380 (0.012)  \\ 
SDSS J152839.42+011300.1  &  16.168 (0.016)  &  16.547 (0.019)  &  17.037 (0.014)  &  57.195  &  57.200  &  57.266  &  57.233 (0.033)  \\ 
SDSS J160839.07+074542.5  &  16.653 (0.019)  &  16.690 (0.016)  &  17.093 (0.016)  &  56.290  &  56.199  &  56.236  &  56.218 (0.019)  \\ 
SDSS J165318.76+371027.2  &  16.563 (0.019)  &  16.498 (0.012)  &  16.843 (0.013)  &  55.954  &  55.847  &  55.826  &  55.837 (0.011)  \\ 
SDSS J165851.11+341853.3  &  15.774 (0.011)  &  16.173 (0.015)  &  16.697 (0.012)  &  57.315  &  57.240  &  57.227  &  57.234 (0.007)  \\ 
SDSS J170331.62+223251.3  &  16.496 (0.016)  &  16.608 (0.011)  &  17.002 (0.013)  &  56.106  &  56.109  &  56.158  &  56.134 (0.025)  \\ 
SDSS J172406.14+562003.1  &  15.830 (0.021)  &  16.029 (0.022)  &  16.419 (0.014)  &  56.741  &  56.557  &  56.438  &  56.497 (0.059)  \\ 
SDSS J212412.14+110415.7  &  16.835 (0.022)  &  16.792 (0.013)  &  17.165 (0.012)  &  56.424  &  56.295  &  56.276  &  56.286 (0.010)  \\ 
SDSS J214001.05-075052.2  &  15.756 (0.020)  &  16.019 (0.012)  &  16.460 (0.013)  &  56.212  &  56.174  &  56.164  &  56.169 (0.005)  \\ 
\hline
\end{tabular}
\end{minipage}
\label{t2}
\end{table*}


\section{Absolute Fluxes}
\label{absolute}

Once the atmospheric parameters, surface gravity and effective temperature, 
have been determined for the DA, their surface radiative flux is given 
by the corresponding model atmospheres.
Only a small part of their spectrum ($385-540$ nm) is actually used in the 
parameter determination described above, but the employed grid of model
fluxes extends to the entire optical region ($300-700$ nm),
and therefore we determine absolute fluxes for the broader spectral region.

\begin{figure*}
\includegraphics[angle=0,width=8cm]{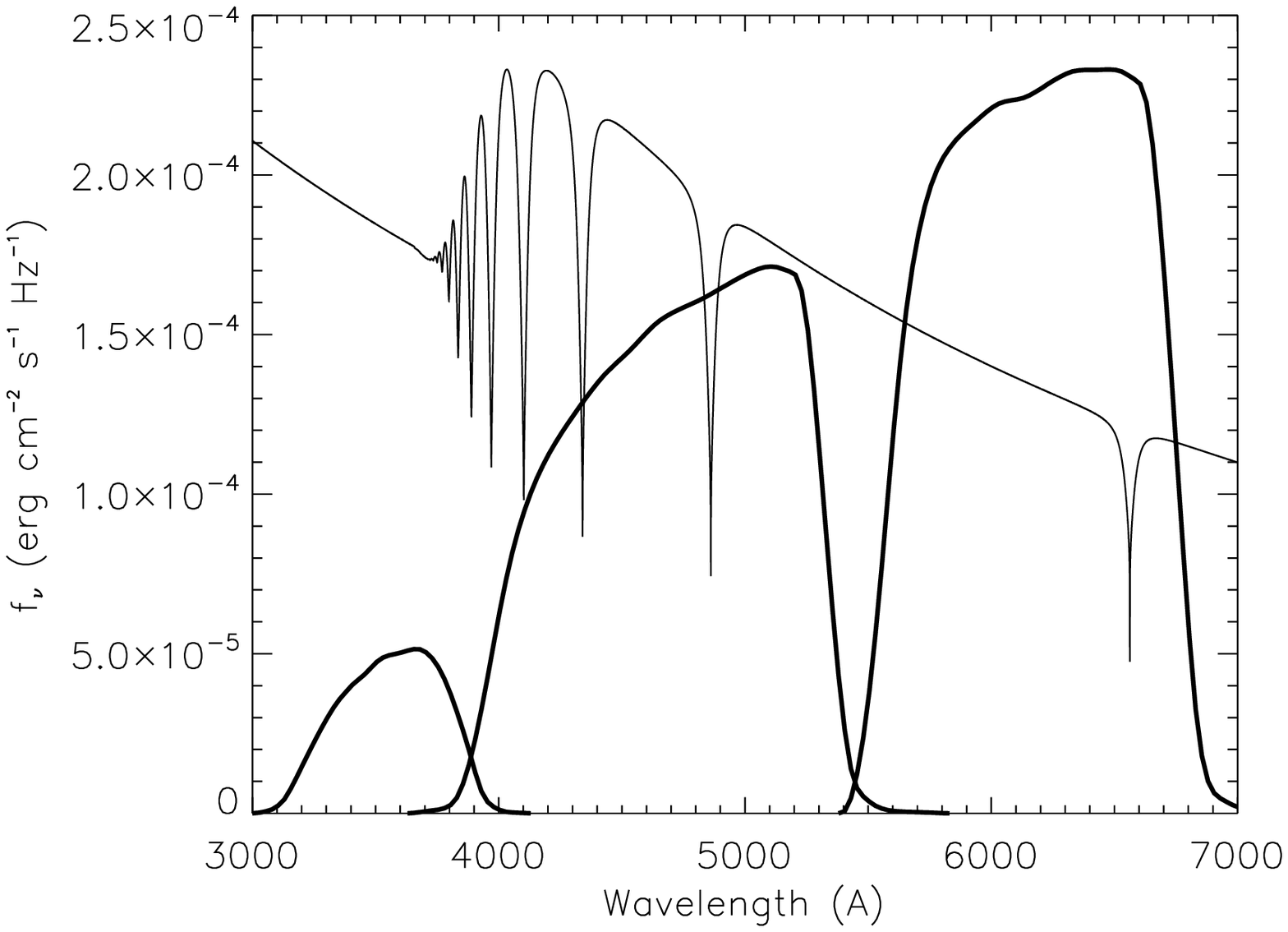}
\caption{Model flux for a DA white dwarf with $T_{\rm eff}=20,000$ and $\log g=7.0$. 
The shapes of the SDSS photometric passbands ($ugr$) are shown with thick lines, 
arbitrarily scaled.}
\label{f5}
\end{figure*}

The quantity of interest is the flux received at the Earth, which is related
to the stellar surface flux provided by the model atmosphere by the angular
diameter, or more precisely by the square of the ratio of the  distance between the star
and the Earth and the stellar radius. Fairly accurate estimates for the radii of
DA white dwarfs are possible based on models, but no direct estimates are
available for the distances to these faint DA. A direct approach that can be followed 
instead is to use broad-band photometry to scale the theoretical
flux distributions. Fig. \ref{f5} illustrates the span of the SDSS $ugr$ bands
and the coverage of the theoretical spectra. With a typical uncertainty
of 0.02 mag in each filter, considering the three together gives an
expected uncertainty in the absolute scale of the fluxes of about 1\%.
This figure is comparable or smaller than the expected uncertainty in the
relative fluxes, but it presumes no systematics.

In practice, we derive synthetic absolute fluxes at the stellar surface for
the corresponding atmospheric parameters and extinction for each DA by
interpolation in our grid, and determine synthetic photometry in the $ugr$ bands
using the filter responses for point sources observed at an airmass of 1.3 
available from the SDSS web 
pages\footnote{http://www.sdss.org/dr3/instruments/imager/index.html\#filters}.
For clarity, we refer to the observed magnitudes with letters in italic
($ugr$, indicating SDSS PSF magnitudes as extracted from the DR6 CAS) 
and to those calculated at the stellar surface in bold 
({\bf ugr}). 
For each star, the difference between the magnitudes computed for the 
the photosphere and those measured at the Earth can be derived for each band as
$\delta m \equiv m-{\bf{m}}$, with $m$({\bf m}) replaced by $u$, $g$ or $r$ 
({\bf u}, {\bf g} or {\bf r}).  
 In the absence of systematic errors, these
three values will be consistent and 
\begin{equation}
 \delta m \equiv m - {\bf m} = -2.5 \log \left(\frac{R}{d}\right)^2,
\end{equation}
\noindent where $R$ is the stellar radius, and $d$ the distance to the star.
(Recall that the effect of interstellar extinction is already included in the
model magnitudes $m$).

Table 2 gives the observed magnitudes and
the $\delta$ values for each star. There is an average offset 
$<\delta u - \delta g> = 0.0590 \pm 0.0056$ mag, 
but the results are more consistent for $g$ and $r$ with 
$<\delta r - \delta g> = -0.0068 \pm 0.0051$ mag. 
The observed $u$ magnitudes should  be reduced by 0.04 magnitudes to
place them on the AB system (Oke \& Gunn 1983), as recommended in the SDSS 
web pages (www.sdss.org; 
see also the discussions in Eisenstein et al. 2006, Ivezi\'c et al. 2007). 
However, after applying this correction, the values for $\delta u$ we derive 
are still offset from those for $g$ and $r$ by 0.02 mag. 
There is no dramatic
variation in the uncertainties in the observations for each filter,
with median values of 0.0179, 0.0176, and 0.0150 mag for $u$, $g$, and $r$,
respectively, but dropping $u$ seems the most sensible option. There
are known issues with the $u$ filters, such as flatfielding problems, 
a red leak which
might affect the observations particularly if there are nearby
unresolved cooler objects (Adelman-McCarthy et al. 2008, Stoughton et al. 2002),  
and a recently discovered variation with time
of the  effective filter transmission caused, most likely, 
by a degradation of the UV
coating of the $u$-band CCDs (Abazajian et al. 2009). In addition,  even 
if the detected $\delta u - \delta g$ offset is removed, the median $1\sigma$
scatter in the average $\delta$ is 0.025 mag, but this value is reduced to 0.016 mag when
only $g$ and $r$ are considered.

\section{The proposed standards}

As discussed in the introduction, the uncertainties in the relative shape of
the spectral fluxes computed for DA stars are at the level of $<1$ \%, but 
given that we do not know exactly the atmospheric parameters and the amount
of interstellar reddening, additional contributions to the error budget 
need to be considered.

\begin{figure}
\resizebox{\hsize}{!}{\includegraphics[angle=90]{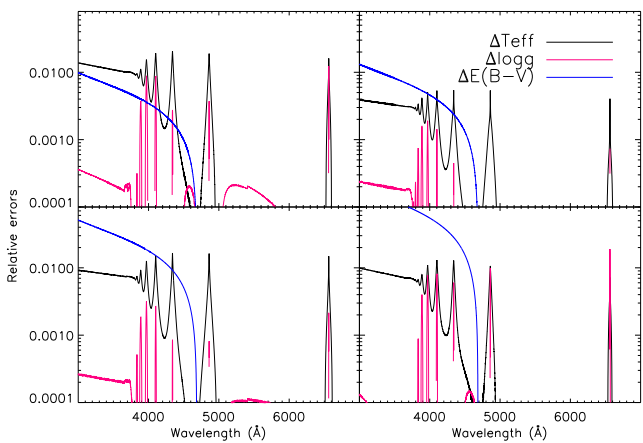}}
\caption{Estimated fractional errors in the relative fluxes for 
the four stars in Fig. \ref{f2}.}
\label{f6}
\end{figure}

To evaluate the impact of the uncertainties in the derived $T_{\rm eff}$,
$\log g$ and $E(B-V)$ on the optical spectral energy distributions assigned
to each of the DA, we interpolate fluxes after perturbing each parameter
by the expected $1-\sigma$ uncertainty. Because we are normalizing the
fluxes of our sample stars to match the SDSS $g$ magnitudes, we normalize
the spectra at the weighted average wavelength for the $g$ passband, which is 4686 \AA.
Fig. \ref{f6} illustrates the results for the four stars included in Fig. \ref{f2}.
Depending on the star, the uncertainties in $T_{\rm eff}$ or $E(B-V)$ dominate 
the errors in the relative fluxes; $\log g$ has always a minor effect.
Large errors are observed in the cores of the Balmer lines, which are particularly
sensitive to changes in $T_{\rm eff}$ and surface gravity.
We combine the three contributions in quadrature to estimate the final uncertainties.
Although it is possible to determine and use the internal covariances among
the errors in the parameters to estimate the uncertainties 
in the fluxes, the  fact is that the errors in the parameters are dominated by
the contributions from the systematic offsets between the analyses using 
continuum-corrected spectra and those in which the continuum shape is preserved,
and therefore we simply assume that the errors are weakly correlated.

\begin{table*}
\label{t3}
 \centering
\begin{minipage}{140mm}
  \caption{Proposed spectrophotometric standards and expected fractional errors in their fluxes.}
  \begin{tabular}{llllll}
  \hline
   Name     &   $g$   &  $\sigma_{\lambda = 3500}$   & $\sigma_{\lambda = 4500}$ & $\sigma_{\lambda = 5500}$ & $\sigma_{\lambda = 6500}$  \\
\hline
SDSS J082346.15+245345.7 & 15.552 & 0.016 & 0.011 & 0.013 & 0.018  \\
SDSS J094203.19+544630.2 & 16.934 & 0.015 & 0.010 & 0.010 & 0.010  \\
SDSS J095230.44+114202.3 & 16.472 & 0.028 & 0.026 & 0.026 & 0.027  \\
SDSS J132434.39+072525.3 & 16.581 & 0.016 & 0.015 & 0.015 & 0.016  \\
SDSS J143315.92+252853.1 & 16.772 & 0.020 & 0.015 & 0.016 & 0.020  \\
SDSS J145415.84+551152.3 & 15.813 & 0.014 & 0.011 & 0.011 & 0.011  \\
SDSS J145600.81+574150.8 & 16.191 & 0.012 & 0.010 & 0.010 & 0.010  \\
SDSS J151421.26+004752.8 & 15.687 & 0.013 & 0.010 & 0.012 & 0.014  \\
SDSS J212412.14+110415.7 & 16.792 & 0.020 & 0.014 & 0.016 & 0.020  \\
\hline
\end{tabular}
\end{minipage}
\end{table*}

The error budget for each of the DA in our sample combines the wavelength-dependent
uncertainties in the relative fluxes, just discussed, with those in the 
zero point magnitudes --set by the average between the observed and predicted 
fluxes in the $g$ and $r$ bands. Useful flux standards should have 
reasonable uncertainties. We select only DA stars
with an total uncertainty $<5$\% at all wavelengths in the optical, and $<3\%$ between
$4500<\lambda<7000$ \AA. There are 9 stars in our sample that satisfy these criteria.
These are compiled in Table \ref{t3}, where we also provide uncertainty
estimates at four different wavelengths. We include 
the expected errors in the theoretical spectral energy distributions  (1 \%).
The theoretical spectral fluxes at the Earth for these stars 
are given in electronic format\footnote{These are available from
\tt http://hebe.as.utexas.edu/std/}.

\section{The HST Standards}

\begin{figure*}
\resizebox{\hsize}{!}{\includegraphics[angle=90]{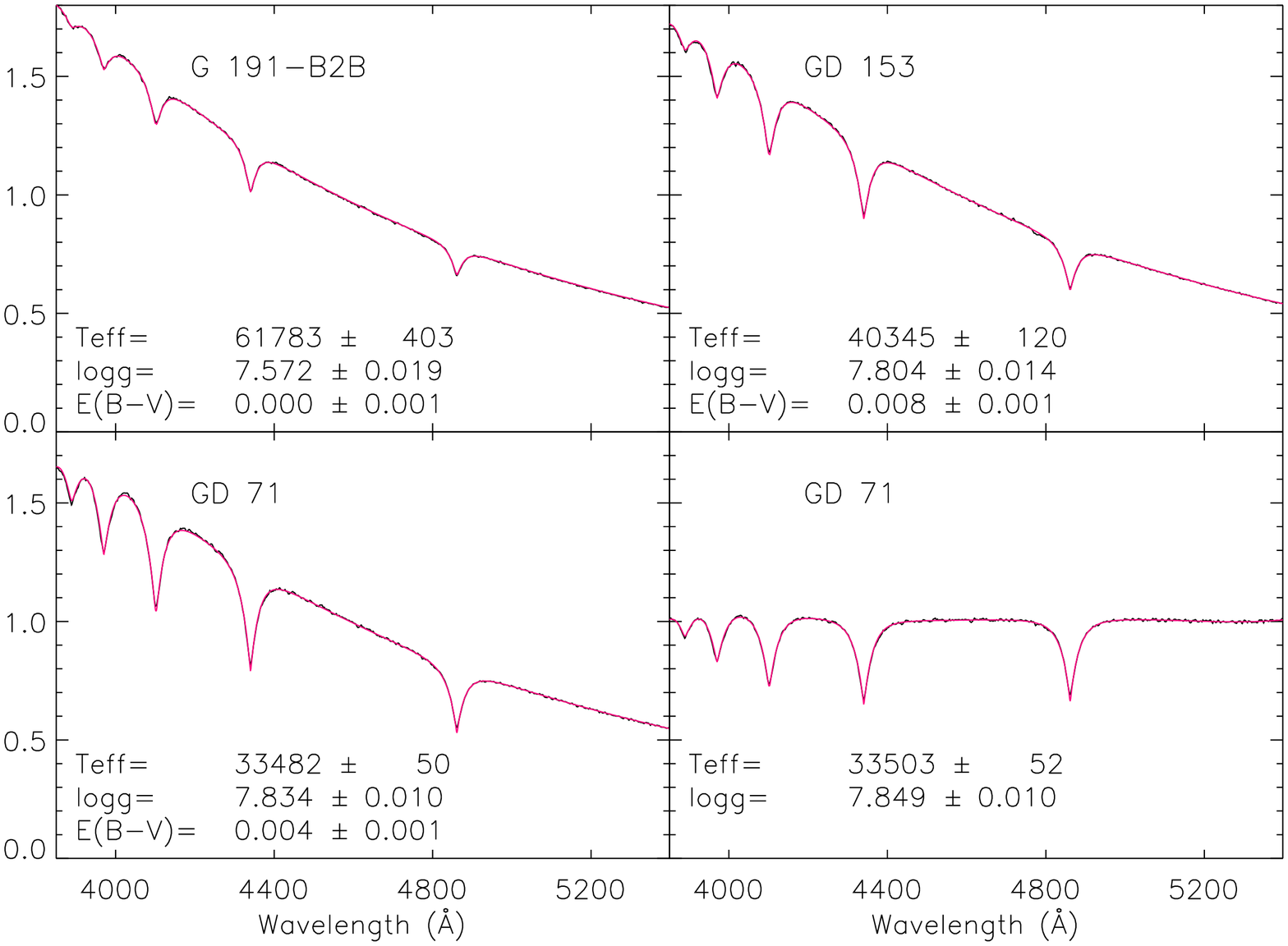}}
\caption{Model fittings to the three HST primary standards. 
The fluxes are in $f_{\lambda}$ units, 
normalized to have a median value of one in the range 3850--5400 \AA
in the top and lower-left panels. The lower-right panel illustrates the 
model fittings for GD 71 in the case when the continuum shape is rectified.
Note that the parameters
here do not match those in Table \ref{t4}, as the table gives the average results 
for two types of analyses: those using spectra that preserve continuum information
 such as those illustrated in the top and the lower-left panels, 
and those using continuum-corrected spectra such as the example for GD 71
in the lower-right panel. The use
of pure-H models for G 191-B2B is inappropriate as a number of heavy metals have
been detected in the UV spectrum of this star.}
\label{f7}
\end{figure*}

As a result of the analysis presented in the previous sections, we have identified
a number of stars for which the absolute fluxes have an expected uncertainty $<2$ \% 
throughout the optical (see Table 3). An obvious question is how these fare in comparison with
the HST standards. To answer this question we follow exactly the same steps 
as for the SDSS stars: fitting the STIS spectra 
normalized to their median in the spectral window, or normalized to the continuum,
adopting averaged $T_{\rm eff}$/$\log g$, etc. 
The original spectra, obtained from the CALSPEC web site,
have a resolving power about 1000 in our range of interest (STIS L-mode 
observations), which is  half of the resolution of the SDSS spectra considered
earlier, and this taken into account in the analysis. 
We first assumed that the spectra were at rest, and after finding the best-fitting
models, we determined Doppler shifts of $+5$, $-18$, and $13 \pm 2$ km 
s$^{-1}$ by cross-correlation, corrected them, and repeated the optimizations.
We refer the reader to Bohlin, Dickinson \& Calzetti
(2001, and references therein) for more details about the observations.
The derived parameters are given in Table 4.
Fig. \ref{f7} is the equivalent of Fig. \ref{f2}  for the HST standards,
and includes the fitting to the continuum-rectified spectrum of GD 71.
Note that the parameters previously used for assigning model fluxes to these stars 
were based on LTE model atmospheres. 

As a reference, we passed the HST fluxes for the three standards through the
SDSS filter responses for $ugr$. The results are
included in Table 5. Holberg \& Bergeron performed the same exercise
and our results agree with theirs to within 0.003 mag, with the exception of
the $g$ band for GD 71, which they found 0.012 mag brighter. As can be
seen in the table, the agreement between the derived $\delta m$ values
across the three bands is better than 0.001 mag for G 191 B2B and
GD 71, and only slightly larger (0.005 mag) for  GD 153, indicating
that the good agreement illustrated in Fig. \ref{f7} extends 
to the  $u$ and $r$ bands.

\begin{table}
\centering
\begin{minipage}{140mm}
  \caption{Atmospheric parameters for the three HST
 standards.}
  \begin{tabular}{llll}
  \hline
   Name     &   $T_{\rm eff}$ & $\log g$ & E(B-V) \\
            &      (K)        &  (dex)   &  (mag) \\
\hline
G 191 B2B   & 61980 (  514) &  7.555 (0.042) &  0.000 (0.001)  \\
 GD 153     & 40401 (  142) &  7.812 (0.018) &  0.008 (0.001)  \\
GD  71      & 33492 (   41) &  7.841 (0.017) &  0.004 (0.001)  \\
\hline
\end{tabular}
\end{minipage}
\label{t4}
\end{table}

\begin{table*}
\centering
\begin{minipage}{140mm}
  \caption{Photometry and zero points for the three HST standards. The uncertainties quoted correspond to the standard error of the mean from the multiple observations available. The PT photometry includes 7 measurements for G 191-B2B, 1 for GD 153, and 5
in $u$ and 6 in $g$ and $r$ for GD 71. The more recent USNO photometry includes 
8 measurements in $u$ and 9 in $g$ and $r$ for G 191-B2B, 4 in all bands for
GD 153, and 12 in $u$ and $r$, and 11 in $g$ for GD 71.}
\begin{tabular}{llllllll}
  \hline
 Name & $u$ & $g$ & $r$ &   $\delta u$ &  $\delta g$ & $\delta r$ & $\delta$ ($g$ and $r$)  \\
\hline
\multicolumn{8}{c}{ HST spectra} \\
\hline
G 191 B2B & 10.990         &   11.467       &   12.008       & 52.597 & 52.594 & 52.595 & 52.594 ($<$0.001) \\
GD 153    & 12.661         &   13.065       &   13.578       & 53.711 & 53.704 & 53.713 & 53.709 (0.005)    \\
GD 71     & 12.422         &   12.772       &   13.259       & 53.143 & 53.135 & 53.136 & 53.136 ($<$0.001)  \\
\hline
\hline
\multicolumn{8}{c}{ PT photometry  (see Holberg \& Bergeron 2006)} \\
\hline
G 191 B2B & 11.033 (0.006) & 11.470 (0.002) & 12.007 (0.003) & 52.639 & 52.597 & 52.593 & 52.595 (0.002) \\
GD 153    & 12.700         & 13.022         & 13.573         & 53.750 & 53.661 & 53.709 & 53.685 (0.024) \\
GD 71     & 12.438 (0.008) & 12.752 (0.005) & 13.241 (0.005) & 53.159 & 53.115 & 53.117 & 53.116 (0.001) \\
\hline
\hline
\multicolumn{8}{c}{ USNO 1-m photometry (see Appendix)} \\
\hline
G 191-B2B & 11.018 (0.004) & 11.457 (0.002) & 12.013 (0.003) & 52.625 & 52.584 & 52.600 & 52.592 (0.008) \\
GD 153    & 12.700 (0.002) & 13.064 (0.002) & 13.592 (0.001) & 53.750 & 53.704 & 53.727 & 53.715 (0.012) \\
GD 71     & 12.442 (0.004) & 12.755 (0.003) & 13.259 (0.002) & 53.163 & 53.118 & 53.136 & 53.127 (0.009) \\
\hline
\end{tabular}
\end{minipage}
\label{t5}
\end{table*}

To set the zero point of the fluxes we examine first the  $ugr$ photometry 
obtained with the Apache Point 0.5-m photometric telescope (PT) 
and transformed to the  2.5-m SDSS scale (obtained by S. Kent and D. Tucker; 
published by Holberg \& Bergeron 2006). The results are included in Table \ref{t5}.
We note, as Holberg \& Bergeron did, an anomalous $g$-band magnitude for GD 153.
After reducing the $u$ magnitudes by 0.04 mag to place them in the AB system, 
the zero points for G1912B2 and GD 71 agree among the 
three bands $ugr$ to about 0.005 mag.
The zero point in GD 153 agrees to 0.001 mag between the $u$ and $r$ bands
(again, after reducing $u$ by 0.04 mag), but these
disagree with that from the $g$ band by 0.05 mag.  
Comparing the differences between the synthetic photometry and the observed magnitudes,
the $\delta$ values, with
those determined for the HST spectra, we find an excellent agreement for G 191 B2B and 
GD 153 (ignoring the $g$ band for the latter). There is, however, a systematic offset 
of about 0.02 mag between the HST absolute fluxes for GD 71 and the 
PT photometry for this star.

Given these issues, we consider new photometry obtained using the USNO
1-m telescope. The new observations are described in the Appendix,
and the mean $ugr$ magnitudes, in the SDSS 2.5-m system, are 
summarized in Table 5. We prefer the new photometry and 
will adopt it in the discussion below.
The formal uncertainties for the USNO photometry ($\sigma/N$; as given in Table 5)  
are typically smaller than for the PT photometry\footnote{Unlike Holberg \& Bergeron
(2006), we quote standard errors of the mean in our table, computed from
the original table kindly provided by D. Tucker.}, due to a larger number
of measurements available.
In addition, while two transformations are required
to place the PT photometry on the SDSS 2.5-m $ugriz$ system, only
one transformation is necessary for the USNO $u'g'r'i'z'$ magnitudes
(Tucker et al. 2006). 
Overall, the
scatter between the $g$ and $r$ band scaling factors ($\delta$) 
is increased relative 
to the earlier PT photometry, but more in line 
with expectations at about 1 \% (Smith et al. 2002). 
The newer observations bring the
$g$-band magnitude of GD 153 in line with the scaling factors determined
for $u$ and $r$. More 
importantly, the 2 \% offset between the GD 71 HST fluxes 
(i.e. the Johnson $V$ magnitude) and 
the PT photometry is now reduced to less then 1\%, and therefore
within the expected variance. 

For the new USNO photometry we find average offsets 
$<\delta u - \delta g > =  0.0442 \pm 0.0017$ mag and 
$<\delta r - \delta g > = +0.0189 \pm 0.0025$ mag. Both of these 
differ by $+0.02$ mag from the offsets we determined in \S 4 
for the SDSS DA stars. Because those in \S 4 correspond to 
measurements made with the 2.5-m telescope, we consider them more 
appropriate for the true SDSS system. 

The uncertainties in the relative
fluxes of these stars due to the uncertainties in their atmospheric parameters and
reddening are smaller than the typical values for the SDSS DA stars.
This is the result of a much higher signal-to-noise ratio of the STIS spectra, 
and the fact that the HST standards are  
brighter and therefore located closer to the Sun and 
affected by essentially no reddening.
Nonetheless, four of the stars in Table \ref{t3} have similar uncertainties as the HST standards
(SDSS  J151421.26+004752.8, SDSS  J094203.19+544630.2 and SDSS  J145600.81+574150.8).
The uncertainties in the absolute fluxes of these
stars are dominated by the theoretical errors, which
limit their accuracy at about 1\%. 

Our inferred surface gravities for the three DA standards are in good agreement with
the values reported by Barstow et al. (2001, 2003a) from the analysis of Lyman and Balmer line
profiles, while our effective temperatures
for GD 71 and GD 153 are a few percent higher than those derived in that work. 
Our temperature for G 191 B2B, however,  is about 15 \% higher than those given by Barstow
et al. from the analysis of Balmer lines, which range between 51500 and 
54500 K\footnote{We note that the Barstow et al. (2003a) 
analysis of FUSE spectra for G 191-B2B leads to higher temperatures (about 57000--59000 K), 
but the differences are believed to be related to 
difficulties modeling the Lyman series.}.
The difference between our analysis and the Balmer-based temperatures compiled by
Barstow et al. can be attributed to the presence of heavy metals  in the atmosphere of 
this star (see, e.g. Lanz et al. 1999).  Our use of a pure-H model leads to an overestimated 
effective temperature (Barstow, Hubeny \& Holberg 1998) by about 6000 K. No metals have been 
detected in the UV spectrum of GD 153 (Barstow et al. 2003b),
 but we have not found a similar study including GD 71.

\begin{figure}
\resizebox{\hsize}{!}{\includegraphics[angle=90]{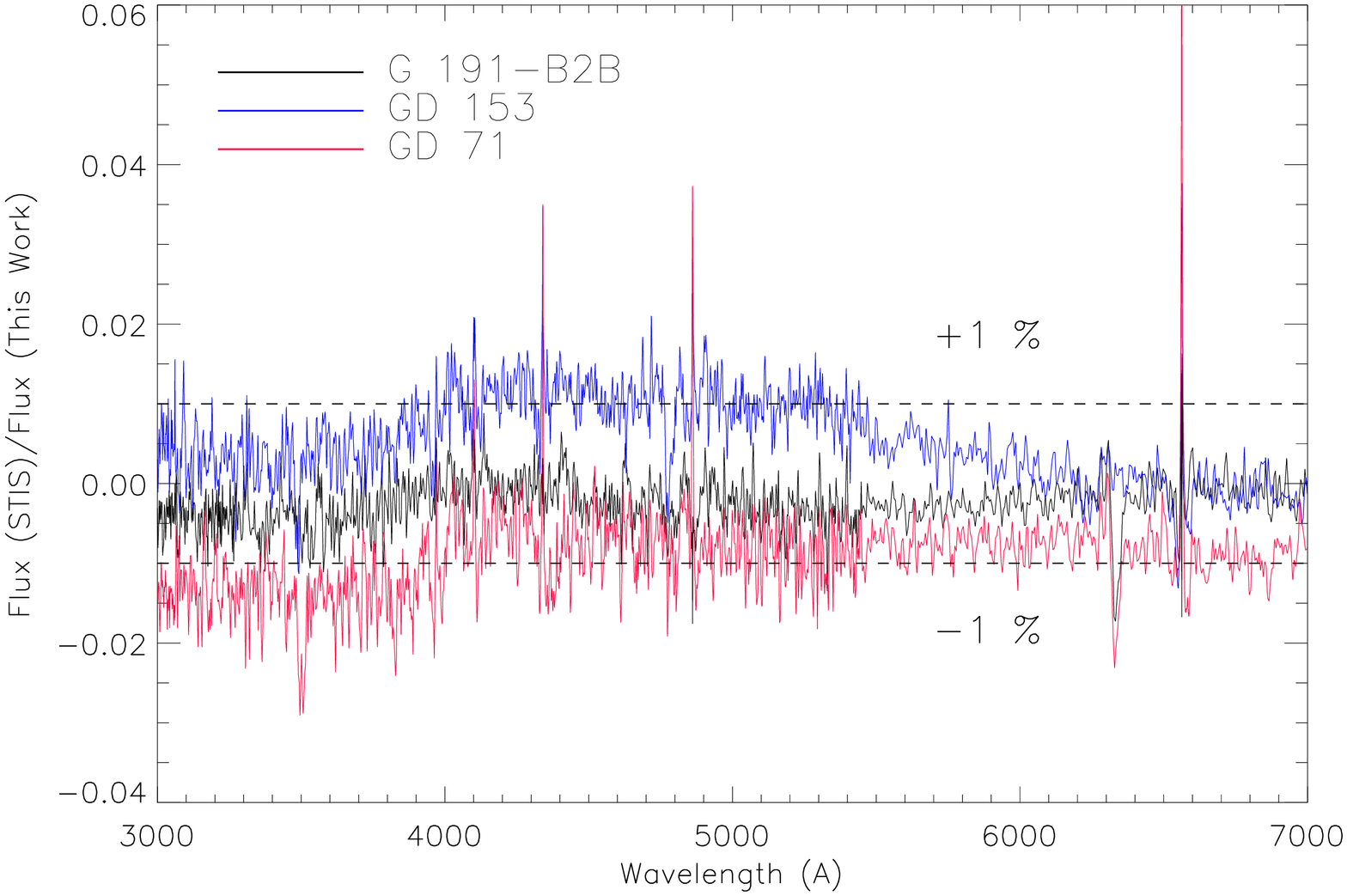}}
\caption{Ratio  between the CALSPEC fluxes and those calculated here using 
redetermined atmospheric parameters for the HST primary standards. 
The zero point of the flux scale of the CALSPEC fluxes is given by 
Landolt $V$-band photometry, while  here it is set 
by SDSS PT $ugr$ photometry (but see text for an exception for GD 153). The two
scales are inconsistent by 1.5 \% for GD 71.}
\label{f8}
\end{figure}

In Fig. \ref{f8} we confront our absolute fluxes for the these stars with those in the 
CALSPEC library. There is good agreement between the spectral shapes, with
the differences caused mainly by the different effective temperatures used 
to these stars by Bohlin (2000; traceable to Finley, Koester \& Basri 1997) 
-- our values are higher by 1, 4 and 2 \% for
G 191-B2B, GD 153 ad GD 71, respectively.  As explained above, our inferred
effective temperature for G191-B2B is overestimated by about  6000 K, but
it is close to that adopted for calibration of the HST fluxes of this star, and hence
the {\it apparent} good agreement shown in Fig. \ref{f8}.
The consistency between the HST and our fluxes is best in the red, but 
overall, the discrepancies are limited to $<1.5$ \%, in line with our 
expectations.

Close inspection of the residuals in Fig. \ref{f8}, ignoring G 191-B2B,
suggests that there is a systematic offset between the $u$ (3000--4000 \AA) 
and $g$-band (4000--5500 \AA) fluxes. The HST fluxes in the $u$ band are
lower than those presented here by about 1 \%. The most likely explanation 
for this feature is the fact that the effective temperatures adopted
in the HST calibration for these stars are lower by 2--4 \%.

\section{Conclusions}

In this paper we consider the possibility of building an extended network of
spectrophotometric standard stars by identifying well-behaved pure-hydrogen
white dwarfs, spectroscopically determining their atmospheric parameters and reddening,
and assigning them absolute fluxes based on model atmosphere calculations scaled with
broad-band photometric observations.
This procedure was successfully used for defining the three  
primary flux standards
for the HST (Finley, Basri \& Bowyer 1990; Bohlin, Colina \& Finley 1995; Bohlin 2000), 
and could be now applied more extensively to many other DA stars that are
being identified in ongoing and planned spectroscopic surveys.

We put this idea to work using publicly available spectra from the Sloan
Digital Sky Survey. We identify 59 spectra of 57 stars that are nicely 
reproduced by NLTE DA models, and measured their atmospheric parameters. 
By comparing the  effective temperatures derived from spectra that have been
continuum rectified and those which preserve the overall spectral shape
we see the effects of reddening in the hotter (and therefore more distant)
stars. By including reddening in the analysis the two sets of temperatures
are brought into agreement.

Our exercise identifies a set of nine new standard candidates with estimated
flux uncertainties less than 3\% between 4500 and 7000 \AA.
We present these stars as {\it potentially} useful standard sources. 
Their observed spectral energy distributions are nicely matched with
DA white dwarf models across the visible, and although we have not verified that 
their fluxes are stable, all the sources are significantly
warmer than the ZZ Ceti window ($\sim 11,000$ K), and therefore it is unlikely that
they show measurable pulsations. Nevertheless, we should exercise caution if
trying to use them as standards at shorter or, especially, longer wavelengths.
Some of these stars could have low-mass stellar or sub-stellar 
companions which could make their fluxes at longer wavelengths to deviate
from the predictions from the stellar models. There could also be 
unresolved nearby objects with variable fluxes whose presence went
unnoticed in the observations analyzed here.

From the comparison of the scaling factors between synthetic and observed 
photometry for the 
full DA sample we find the following offsets:
$<\delta u - \delta g> = 0.059 \pm 0.006$ mag and 
$<\delta g - \delta r> = -0.007 \pm 0.005$ mag. Assuming the $g$ and the
$r$-bands correspond exactly to AB magnitudes (see, e.g., the discussion 
by Eisenstein et al. 2006), and that the pure-H models provide accurate descriptions 
of the spectral energy distribution of the DA stars, the derived $\delta u - \delta g$
difference could simply correspond to the AB mag correction for the $u$ band.
However, we note that Holberg \& Bergeron  (2006) found indication of
an offset of $-0.02$ mag 
between the SDSS magnitudes for 107 DA white dwarfs in the $g$ band
and their fluxes in the Johnson $V$ band. 

We carry out a reanalysis of the STIS spectra for the three primary
HST standards GD 71, GD 153 and G 191-B2B.
We use NLTE model atmospheres to determine their
atmospheric parameters, and find effective temperatures that are hotter
by 1--4 \% compared to those adopted based on an LTE analysis. 
Our predicted fluxes for the three standards, scaled using $ugriz$ photometry,
agree  at a $1$ \% level with the adopted HST fluxes.

Heavy metals are present in the atmosphere of G 191-B2B, and our 
analysis based on pure-H models results in an effective temperature
overestimated by about 6000 K. The good agreement for GD 153 and GD 71 
demonstrates consistency, but that found for G 191-B2B  is simply 
an artifact and highlights that the spectral energy distribution 
adopted for  this standard  needs to be revised.
Our analysis also unveils an inconsistency between the photometry
obtained for GD 71 with the Apache Point 0.5-m Photometric Telescope, and the zero
point of the STIS fluxes for this star,  which is  based on 
Landolt (1992) V-band photometry, but the  discrepancy is greatly reduced when
considering new $u'g'r'$ photometry from the USNO 1-m telescope. Overall,
our study supports the level of accuracy presumed for the HST standards,
but we also find room for improvement at the 1 \% level.

\section*{Acknowledgments}

We are thankful to Douglas Tucker for comments about the uncertainties of 
the PT photometry, and to Ralph Bohlin for useful discussions about
HST spectrophotometry. Jay Holberg provided a number of stimulating 
comments which improved this paper.


\appendix

\section{USNO $u'g'r'i'z'$ photometry for the HST standards}

The observations  were obtained using  the 1.0-m Ritchey--Chr\'{e}tien
telescope at the  U.S. Naval Observatory's Flagstaff Station (USNO-FS)
during the   bright time between  October 2005 and April 2006.
The detector was a thinned, UV-AR
coated, Tektronix TK1024 CCD with a gain of 7.43$\pm$0.41 e$^-$
and a read-noise  of 6.0 e$^-$.  This CCD is
similar to the CCDs used in the main SDSS  survey camera and the CCD
used by the 0.5-m photometric monitoring telescope at APO.  The camera
scale of 0.68 arcsec/pixel produced a  field of view of 11.54 arcmin.
  
The five filters of the $u'g'r'i'z'$ system have effective wavelengths
of 3540 \AA, 4750 \AA, 6222 \AA, 7632 \AA, and 9049 \AA, respectively,
at 1.2 airmasses (Fukugita et al. 1996; note that the $g'$ filter has been
determined to have an effective wavelength 20 \AA\, bluer than that
originally quoted).
The filter transmission data as measured by the Japan Participation
Group within the SDSS, and the
CCD+filter response curves updated by J. Gunn in 2001 are available 
on-line\footnote{\tt http://www-star.fnal.gov/ugriz/Filters/response.html}. 
We refer the reader to Smith et al. 2002 for more details about 
the observations,  and to Smith et al. 2007 for a historical description
of the SDSS photometric system.

All the individual measurements are given in Table \ref{usno}. 
About  9, 4 and 12 observations were obtained for G191 B2B, GD 153
and GD 71, respectively. The exposure times were (in order $u'g'r'i'z'$)
60, 15, 15, 40, 90 s for G191 B2B, 
100, 25, 25, 60, 130 s for GD 153, and
20, 8, 10, 12, 50 for GD 71. 
The average $u'g'r'i'z'$ magnitudes from the USNO 1.0-m,
included in Table \ref{usno},  were converted
to the SDSS 2.5-m $ugriz$ system using the Equations 2--6  
in Tucker et al. 2006\footnote{Also given at
http://www.sdss.org/dr7/algorithms/jeg\_photometric\_eq\_dr1.html}. 
We note that the transformations were derived from stars cooler than
our standards. Rider et al. (2004) found that they
are still adequate for colors bluer than originally defined, although
still far from the extremely blue colors of the  white dwarfs considered 
in this paper ($u-g < -0.3$).
The derived $ugr$ magnitudes are included in  Table 5.

\begin{table*}
\label{usno}
 \centering
\begin{minipage}{140mm}
  \caption{Photometry for the HST standards in the USNO $u'g'r'i'z'$ bandpasses.}
  \begin{tabular}{llllll}
  \hline
  MJD $\equiv$ JD -240,000,000.5 &       $u'$  &    $g'$  &    $r'$  &    $i'$   &   $z'$     \\
  (days) &   (mag)  &    (mag) &   (mag)  &    (mag)  & (mag)     \\
\hline
         &  \multicolumn{5}{c}{G 191 B2B} \\
\hline
 53650  &  11.006 & 11.517 & 12.034 & 12.419 & 12.739  \\
 53651  &  11.024 & 11.533 & 12.046 & 12.421 & 12.754  \\
 53654  &  11.011 & 11.513 & 12.019 & 12.402 & 12.713  \\
 53685  &  11.028 & 11.525 & 12.049 & 12.425 & 12.739  \\
 53687  &  11.035 & 11.523 & 12.032 & 12.422 & 12.710  \\
 53702  &  11.006 & 11.524 & 12.035 & 12.410 & 12.731  \\
 53714  &  11.017 & 11.511 & 12.025 & 12.403 & 12.735  \\
 53715  &  11.017 & 11.513 & 12.039 & 12.419 & 12.734  \\
 53740  &  \dots  & 11.519 & 12.025 & 12.406 & 12.732  \\
\hline
 Mean   &  11.018 & 11.520 & 12.034 & 12.414 & 12.732 \\
 Std. Dev.& 0.011 &  0.007 &  0.010 &  0.009 &  0.013 \\
 Std. Err.& 0.004 &  0.002 &  0.003 &  0.003 &  0.005 \\
\hline
\hline
        &  \multicolumn{5}{c}{GD 153} \\
\hline
 53797  & 12.695 & 13.132 & 13.615 & 13.974 & 14.304  \\
 53832  & 12.701 & 13.125 & 13.611 & 13.972 & 14.260  \\
 53833  & 12.704 & 13.121 & 13.611 & 13.968 & 14.300  \\
 53850  & 12.700 & 13.123 & 13.610 & 13.982 & 14.290  \\
\hline
 Mean   & 12.700 & 13.125 & 13.612 & 13.974 & 14.289  \\
Std. Dev.& 0.004 &  0.005 &  0.002 &  0.006 &  0.020  \\
Std. Err.& 0.002 &  0.002 &  0.001 &  0.003 &  0.010  \\
\hline
\hline
         &  \multicolumn{5}{c}{GD 71} \\
\hline
 53650  & 12.422 & 12.814 & 13.274 & 13.646 & 13.973   \\
 53650  & 12.426 & \dots  & 13.279 & \dots  & 13.987   \\
 53651  & 12.451 & 12.819 & 13.282 & 13.656 & 13.979   \\
 53654  & 12.438 & 12.809 & 13.271 & 13.634 & 13.962   \\
 53687  & 12.457 & 12.819 & 13.284 & 13.652 & 13.965   \\
 53702  & 12.423 & 12.802 & 13.274 & 13.644 & 13.948   \\
 53714  & 12.443 & 12.799 & 13.274 & 13.637 & 13.967   \\
 53715  & 12.447 & 12.820 & 13.283 & 13.650 & 13.966   \\
 53740  & 12.444 & 12.812 & 13.277 & 13.639 & 13.958   \\
 53742  & 12.462 & 12.822 & 13.295 & 13.667 & 13.963   \\
 53832  & 12.451 & 12.828 & 13.286 & 13.645 & 13.976   \\
 53833  & 12.445 & 12.818 & 13.275 & 13.651 & 13.959   \\ 
 \hline
 Mean   & 12.442 & 12.815 & 13.280 & 13.647 & 13.967  \\
Std. Dev.& 0.013 &  0.009 &  0.007 &  0.009 &  0.011  \\
Std. Err.& 0.004 &  0.003 &  0.002 &  0.003 &  0.003  \\
\hline
\hline
\end{tabular}
\end{minipage}
\end{table*}

\label{lastpage}

\end{document}